\documentclass[preprint2]{aastex}

\usepackage{caption}
\usepackage{hyperref}
\usepackage{natbib}
\usepackage[x11names]{xcolor}
\usepackage[countmax]{subfloat}
\usepackage{subfig}

\begin{document}

\title{Inside catalogs: a comparison of source extraction software}

\author{M. Annunziatella}
\affil{Physics Department, Federico II University, via Cinthia 6,I-80126, Napoli, Italy}

\author{A. Mercurio}
\affil{INAF, Astronomical Observatory of Capodimonte, via Moiariello 16, I-80131Napoli, Italy}

\author{M. Brescia}
\affil{INAF, Astronomical Observatory of Capodimonte, via Moiariello 16, I-80131Napoli, Italy}
\affil{Department of Physics, Federico II University, via Cinthia 6,I-80126, Napoli, Italy}

\author{S. Cavuoti}
\affil{Department of Physics, Federico II University, via Cinthia 6,I-80126, Napoli, Italy}

\author{G. Longo}
\affil{Department of Physics, Federico II University, via Cinthia 6,I-80126, Napoli, Italy}
\affil{Visiting associate- Department of Astronomy, California Institute of Technology, CA 90125, USA}

\begin{abstract}

The scope of this paper is to compare the catalog extraction performances obtained using the new combination of SExtractor with PSFEx, against the more traditional and diffuse application of DAOPHOT with ALLSTAR; therefore, the paper may provide a guide for the selection of the most suitable catalog extraction software. Both software packages were tested on two kinds of simulated images having, respectively, a uniform spatial distribution of sources and an overdensity in the center. In both cases, SExtractor is able to generate a deeper catalog than DAOPHOT. Moreover, the use of neural networks for object classification plus the novel {\tt{SPREAD\_MODEL}} parameter push down to the limiting magnitude the possibility of star/galaxy separation. DAOPHOT and ALLSTAR provide an optimal solution for point-source photometry in stellar fields and very accurate and reliable PSF photometry, with robust star-galaxy separation. However, they are not useful for galaxy characterization, and do not generate catalogs that are very complete for faint sources. On the other hand, SExtractor, along with the new capability to derive PSF photometry, turns to be competitive and returns accurate photometry also for galaxies. We can assess that the new version of SExtractor, used in conjunction with PSFEx, represents a very powerful software package for source extraction with performances comparable to those of DAOPHOT. Finally, by comparing the results obtained in the case of a uniform and of an overdense spatial distribution of stars, we notice, for both software packages, a decline for the latter case in the quality of the results produced in terms of magnitudes and centroids.\\

\end{abstract}


\graphicspath{{./}}

\section{Introduction}

Over the past two decades, advances in technology are progressively moving us beyond the traditional observational paradigm, in which most astronomical studies were made by individual observations of small samples of objects. Modern digital sky surveys are already producing increasing amounts of data flows that cannot be effectively handled with traditional methods. This has profoundly changed the needs of scientists in terms of software and data-analysis methods. For instance, the sheer size of the raw data makes it almost impossible to re-process the raw images and, therefore, large catalogs are becoming the primary source of information. \\
When extracting a catalog, the main aspects to take into account are: to detect as many sources as possible; to minimize the contribution of spurious objects; to correctly separate sources in their classes (e.g. star/galaxy classification); to produce accurate measurements of photometric quantities; and, finally, to obtain accurate estimates of the positions of the centroids of the sources.\\
Among the main source extraction software packages used by the astronomical community, there are SExtractor \citep{bertin96} and DAOPHOT II \citep{stetson87}, and the latter is often used in combination with its companion tool ALLSTAR \citep{stetson94}.
SExtractor is commonly used in extragalactic astronomy and has been designed to extract a list of measured properties from images, for both stars and galaxies. DAOPHOT and ALLSTAR were designed to perform mainly stellar photometry.
So far, only DAOPHOT II, used together with ALLSTAR, has been able to produce more accurate photometry for stellar objects using the Point Spread Function (PSF) fitting technique, while the PSF fitting photometry in SExtractor has,
instead, become possible only in the recent years.
The first attempt was in the late 90s, when the PSFEx (PSF Extractor) software package became available within the TERAPIX ``consortium''.
This tool extracts precise models of the PSF from images processed by SExtractor. Only after 2010, through the public release of PSFEx (\citealp{bertin11})\footnote{Available at \url{http://www.astromatic.net/software/psfex}.}, and with the recent evolution of computing power, has PSF fitting photometry become fully available in SExtractor.\\
The scope of this paper is to compare the results obtained using the combination of SExtractor with PSFEx, and DAOPHOT with ALLSTAR, by focusing, in particular, on the completeness and reliability of the extracted catalogs, on the accuracy of photometry, and on the determination of centroids, both with aperture and PSF-fitting photometry.
A previous comparison among extraction software tools was performed by \citet{becker07}. They,
in pursuit of LSST science requirements, performed a comparison among DAOPHOT, two versions of SExtractor (respectively 2.3.2 and 2.4.4), and DoPhot \citep{mateo89}. However, differently from the present work where simulations are used, they evaluated as ``true'' values the measurements obtained with the SDSS imaging pipeline \textit{photo} \citep{lupton01}. Furthermore, we wish to stress that their results were biased by the fact that in 2007 the PSF fitting feature had not yet been implemented in SExtractor.\\
The present work performs, for the first time, a comparison between DAOPHOT and SExtractor PSF photometry, providing a guide for the selection of the most suitable catalog extraction software packages.\\
The simulations used for the comparison are described in Sect.~\ref{2}. In Sect.~\ref{3}, the main input parameters of the software packages are overviewed and the adopted values are specified in Sect.~\ref{4}. In Sect.~\ref{5}, the obtained results are shown. In order to better evaluate the performances of both software packages on crowded fields, we describe: in Sect.~\ref{6}, a test performed on an image showing an overdensity in the center. Finally, the results are summarized in Sect.~\ref{7}, together with our conclusions.\\

\section{Image Simulations}
\label{2}
Image simulations are suitable in testing performances of various software packages. Simulations, in fact, allow us to know exactly the percentage and the type of input sources and their photometric properties. \\
In the present work, simulations have been obtained by using two software packages: Stuff\footnote{Available at \url{http://www.astromatic.net/software/stuff}.} and SkyMaker\footnote{Available at \url{http://www.astromatic.net/software/skymaker}.}, developed by E. Bert\`{\i}n.
With these tools, it is possible to reproduce the real outcome of a CCD observation, once the characteristics of the telescope and the camera are known.
In practice, Stuff can be used to produce a realistic simulated galaxy catalog, while SkyMaker uses this catalog to produce an optical image under realistic observing conditions, allowing also to add a stellar field. \\
The galaxy catalog simulated by Stuff can be produced so to be consistent with the assumed cosmological model and with the statistical distributions of stars and galaxies in terms of redshift, luminosity and color. In a binned redshift space, Stuff produces galaxies of different Hubble types E, S0, Sab, Sbc, Scd and Sdm/Irr. The number of galaxies in each bin is determined from a Poisson distribution, by assuming a non-evolving Schechter luminosity function \citep{schechter76}. Cosmological parameters, luminosity function, as well as instrumental parameters are specified by the user in the input configuration file.
In particular, the size of the image, the pixel scale, the detector gain, and the observed passbands must be provided as input information. Filters can be selected among the many available in the wavelength range [0.29, 87.74831] $\mu$m. Finally the magnitude range of simulated galaxies has to be fixed.\\
The image is then created by SkyMaker, by rendering sources of the input catalog in the frame at the specified pixel coordinates, to which is added a uniform sky background, Poissonian noise, and Gaussian read-out noise. Stellar sources are modeled using a PSF internally generated, while the various types of galaxies are modeled as differently weighted sums of a bulge profile and an exponential disk. The PSF profile takes into account both atmospheric seeing and optical aberrations.
There are many parameters to be set in the SkyMaker configuration file. The most important are related to: the pupil features, e.g., the size of the mirrors and the aberration coefficients; the detector characteristics, e.g., gain, saturation level and image size; and the observing conditions, e.g, full width at half maximum of the seeing (FWHM) and exposure time.\\
For our work, we simulated images as they would be observed by the VST\footnote{\url{http://www.eso.org/public/teles-instr/surveytelescopes/vst/surveys.html}} (VLT Survey Telescope) and the OmegaCAM camera\footnote{\url{http://www.astro-wise.org/~omegacam/index.shtml}}. The field of view (FoV) of OmegaCAM@VST is 1 square degree, with a pixel scale of 0.213 arcsec/pixel. In order to reduce the computational time, we limited our simulations to a FoV of 1/4 of VST.
The aberration coefficients, the tracking errors, and the positions of the spiders were set properly, according to the VST technical specifications.
The allowed magnitude range was set to 14-26 mag and the exposure time was fixed at 1500 s.
Finally, according to the ESO statistics at Cerro Paranal, the FWHM of the seeing was set to 0.7 arcsec.
In this paper, we report results relative to B-band simulated images.
We obtain a catalog of N=4120 sources down to the input magnitude limit. This corresponds to an image with an average surface number density of $\sim$ 20 sources/arcmin$^\mathrm{2}$. \\
In order to perform a more complete comparison of both software packages, we also simulated a crowded image with an overdensity in the center (see Sect.~\ref{6}). To this purpose, the parameters of Stuff and SkyMaker are equal to those used for images with uniform source distribution, except for the image size and the exposure time. For these second group of tests, we reduced the exposure time to 300 s, with an input magnitude limit of 14-25, while the image size is set to 2048x2048 pixels ($\mathrm{\sim}$ 7x7 arcmin$^\mathrm{2}$).
We wish to stress that, in the simulations, we did not include any artifact such as bad pixels, ghosts or bad columns or other effects that, while being crucial in other applications, are of no interest here, since we are comparing the performances of two different software packages. \\

\section{Source extraction software}
\label{3}

In the following section, we briefly discuss how DAOPHOT works in combination with ALLSTAR, and how SExtractor works in combination with PSFEx, and give a brief overview of the main parameters that are needed to be set in order to optimally run the selected software packages.

\subsection{DAOPHOT II}
\label{3.1}

DAOPHOT II is composed of a set of routines mainly designed to perform stellar photometry and astrometry in crowded fields. \\
It requires several input parameters, listed in the file \textit{daophot.opt}, including detector gain, readout noise ({\tt{GAIN, READ NOISE}}), saturation level ({\tt{HIGH GOOD DATUM}}), approximate size of unresolved stellar sources in the frame ({\tt{FITTING RADIUS}}), PSF radius ({\tt{PSF RADIUS}}), PSF model ({\tt{ANALYTIC MODEL PSF}}), and a parameter designed to allow the user to visually inspect the output of each routine ({\tt{WATCH PROGRESS}}).\\
The first step performed by DAOPHOT II is to estimate the sky background and to find the sources above a fixed input threshold through the FIND routine.\\
This threshold represents the level (in ADU), above the sky background, required for a source to be detected. In order to ignore smooth large-scale variations in the background level of the frame, the image is convolved with a lowered truncated Gaussian function, whose FWHM is equal to the input value set by the {\tt{FWHM}} parameter. After the convolution, the program searches for the local maxima sky enhancement. \\
Once the sources are detected, DAOPHOT II performs aperture photometry via the PHOTO routine.
Aperture photometry usually requires the definition of at least two apertures. The first one is usually circular, centered on the source and with a radius of a few times its FWHM. The second one is, instead, ring-shaped; usually this is concentric to the first one and it has an inner radius equal to the radius of the first aperture. The ring-shaped aperture is used to estimate the sky contribution and it usually covers a number of pixels equal to or at least comparable with that of the aperture.
Then, the flux of the source is obtained by subtracting the sky flux from the aperture flux. The aperture size must be chosen thoroughly. In fact, if the radius of the inner aperture is too small, there will be a flux loss; while, if it is too large, too much sky is included and the measurements will become too noisy. The radii of the aperture and sky annulus for DAOPHOT can be specified in an input file: \textit{photo.opt}. The inner and outer radii of the sky annulus, which is centered on the position of each star, must also be specified.\\
Beside the source magnitude, the PHOTO routine produces the coordinates of the source centroids, corresponding to the barycenters of the intensity profile around the source. \\
Aperture photometry performs rather well under the hypothesis of bright and isolated stars. However, the stars in crowded fields are faint and tend to overlap. In these cases, the PSF fitting photometry can produce better results. This measurement requires a PSF model to be derived from the stars in the image of interest. The normalized PSF model is then fitted to each star in the image, in order to obtain the intensity and magnitude. \\
DAOPHOT II can build a PSF model from a sample of stars obtained with the PHOTO routine in an iterative procedure, intended to subtract neighboring stars that might contaminate the profile.
Among them, DAOPHOT will exclude stars within one radius from the edges of the image and the stars too close to saturated stars. The analytical formula of the PSF is chosen by the user among several available models: a Gaussian function, two implementations of a Moffat function, a Lorentz function, and two implementations of a Penny function (\citealp{penny95}). The PSF routine produces a PSF model and a list of the PSF stars and their neighbors. The modeled PSF stars can be visually inspected by setting properly the aforementioned {\tt{WATCH PROGRESS}} parameter. \\
Although DAOPHOT is designed for stellar photometry, extended sources are likely always present in real images and, therefore, a reliable method is required to separate galaxies from stars.\\
The sharpness parameter (\texttt{SHARP}) can be used as star/galaxy classifier. \texttt{SHARP} describes how much broader is the actual profile of the object compared to the profile of the PSF.
The sharpness is, therefore, dependent on the model of the PSF, and can be easily interpreted by plotting it as a function of apparent magnitude. Objects with \texttt{SHARP} significantly greater than zero are probably galaxies.\\
Throughout the paper, we indicate simply with DAOPHOT the stand-alone DAOPHOT II version 1.3-2\footnote{Available at \url{http://starlink.jach.hawaii.edu/starlink}}.

\subsection{ALLSTAR}
\label{3.2}
After having derived PSF models with DAOPHOT, ALLSTAR simultaneously fits multiple overlapping point-spread functions to all the detected sources in the image. At each iteration, ALLSTAR subtracts all the stars from a working copy of the input image, according to the current best guesses about their positions and magnitudes. Then, it computes the increments to the positions and magnitudes by examining the subtraction residuals around each position. Finally, it checks each star to see whether it has converged or it has become insignificant. When a star has converged, its coordinates and magnitude are written in the output file, and the star is permanently subtracted from the working copy of the image; when a star has disappeared, it is simply discarded.\\
The input parameters for ALLSTAR, listed in \textit{allstar.opt}, are similar to those in \textit{daophot.opt} and \textit{photo.opt}.\\
Moreover, it is possible to optimize the determination of centroids, by applying a PSF correction and setting the option {\tt{REDETERMINE CENTROIDS}}.\\
An improvement of the star/galaxy classification is possible by using the aforementioned sharpness measure obtained by ALLSTAR (\texttt{SHARP}). This value may be used also in conjunction with another ALLSTAR output parameter, for instance $\mathrm{\chi}$, which is the observed pixel-to-pixel scatter from the model image profile, divided by the expected pixel-to-pixel scatter from the image profile.
Throughout the paper, we indicate simply with ALLSTAR the PSF-fitting software package that comes together with DAOPHOT II version v. 1.3-2.

\subsection{SExtractor}
\label{3.3}

SExtractor is a software package mainly designed to produce photometric catalogs for a large number of sources, both point-like and extended. Sources are detected in four steps: \textit{i)} sky background modeling and subtracting, \textit{ii)} image filtering, \textit{iii)} thresholding and image segmentation, \textit{iv)} merging and/or splitting of detections. The final catalog is, indeed, extracted according to the input configuration file, in which parameters are set by the user.\\
The first step of the background estimation can be skipped if the user provides manually an input estimation of sky background.
For the automatic background estimation, the most critical input parameters to be set are {\tt{BACK\_SIZE}}, the size of each mesh of the grid used for the determination of the background map, and {\tt{BACK FILTERSIZE}}, the smoothing factor of the background map.\\
Once the sky background is subtracted, the image must be filtered. This implies convolving the signal with a mask, shaped according to the characteristics that the user wants to enhance in the image data.
In fact, there are different filters available in SExtractor. The more suitable are ``top-hat'' functions, optimized to detect extended, low-surface brightness objects, Gaussian functions usually used for faint object detection, and ``Mexhat'' filters, which work with a high value of detection threshold, suitable for bright detections in very crowded star fields.\\
The detection process is mostly controlled by the thresholding parameters ({\tt{DETECT\_THRESHOLD}} and {\tt{ANALYSIS\_THRESHOLD}}). The choice of the threshold must be carefully considered. A too high threshold determines the loss of a high number of sources in the extracted catalog, while a too low value leads to the detection of spurious objects. Hence it is necessary to reach a compromise by setting these parameters according to the image characteristics, the background RMS, and also to the final scientific goal of the analysis.\\
Two or more very close objects can be detected as a unique connected region of pixels above threshold and, in order to correct for this effect, SExtractor adopts a deblending method based on a multi-thresholding process. Each extracted set of connected pixels is re-thresholded at N levels, linearly or exponentially spaced between the initial extraction threshold and the peak value. Also, here a compromise is needed to be found, since a too low value for the deblending parameter leads to a lack of separation between close sources, while a too high value leads to split extended faint sources in more components.
Alternatively, it is possible to extract the catalog with different deblending parameters and to merge detections for extended sources or close pairs.\\
Once sources have been detected and deblended, the software tool starts the measurement phase. SExtractor can produce measurements of position, geometry, and of several types of photometric parameters, including different types of magnitudes.
Among photometric quantities, there are: the aperture magnitude ({\tt{MAG\_APER}}), having the same meaning as explained in Sect.~\ref{3.1}, the Kron magnitude ({\tt{MAG\_AUTO}}), which is the magnitude estimated through an adaptive aperture \citep{kron80}, and the isophotal magnitude ({\tt{MAG\_ISO}}), computed by considering the threshold value as the lowest isophote.\\
Among the available position parameters, it is important to mention the barycenter coordinates ({\tt{X\_IMAGE}}, {\tt{Y\_IMAGE}}), computed as the first order moments of the intensity profile of the image, and windowed positional parameters ({\tt{XWIN\_IMAGE}}, {\tt{YWIN\_IMAGE}}), computed in the same way as the barycenter coordinates, except that the pixel values are integrated within a circular Gaussian window as opposed to the object's isophotal footprint.\\
To separate extended and point-like sources, it is possible to use the \textit{stellarity index} ({\tt{CLASS\_STAR}}), which results from a supervised neural network that is trained to perform a star/galaxy classification. {\tt{CLASS\_STAR}} can assume values between 0 and 1. In theory, SExtractor considers objects with {\tt{CLASS\_STAR}} equal to zero to be galaxies, and those with value 1 as a star. In practice, stars are classified by selecting a {\tt{CLASS\_STAR}} value above 0.9.
Two other parameters, often used to discriminate between star and galaxies, are the half-light radius ({\tt{FLUX\_RADIUS}}), and the peak surface brightness above background (\texttt{$\mathrm{\mu_{max}}$}). When plotted against the Kron magnitude, these two parameters identify a so-called stellar locus.\\
Throughout the paper, we indicate simply with SExtractor the software version 2.14.7 (trunk.r284).

\subsection{PSFEx}
\label{3.4}

The last version of SExtractor can work in combination with PSFEx, a package able to build a model of the image PSF. The latter is expressed as a sum of N$\mathrm{\times}$N pixel components, each one weighted by the appropriate factor in the polynomial expansion (see \citealt{mohr12}). Then, SExtractor takes the PSFEx models as input and uses them to carry out the PSF-corrected model fitting photometry for all sources in the image.\\
PSFEx requires, as input, a catalog produced by SExtractor to build a model of the image PSF, which can be read back in a second run by SExtractor itself. In order to allow PSFEx to work, the first catalog produced by SExtractor must contain at least a given number of parameters, as explained in the PSFEx manual\footnote{\url{https://www.astromatic.net/pubsvn/software/psfex/trunk/doc/psfex.pdf}}.	
In particular, the catalog must contain the parameter {\tt{VIGNET}}, a small stamp centered on each extracted source, used to model the PSF. The size of {\tt{VIGNET}} must be chosen accordingly to the size of the photometric apertures defined by {\tt{PHOT\_APERTURES}}. \\
PSFEx models the PSF as a linear combination of basis vectors. These may be the pixel basis, the Gauss-Laguerre or Karhunen-Lo\`eve bases derived from a set of actual point-source images, or any other user-provided basis. The size of the PSF and the number and type of the basis must be specified in the configuration file.\\
By using SExtractor combined with PSFEx, it is possible to obtain various estimates of the magnitude, in addition to those described in the previous section: the magnitude resulting from the PSF fitting ({\tt{MAG\_PSF}}), the point source total magnitude obtained from fitting ({\tt{MAG\_POINTSOURCE}}), the spheroidal component of the fitting ({\tt{MAG\_SPHEROID}}), the disk component of the fitting ({\tt{MAG\_DISK}}), and the sum of the spheroid and disk components ({\tt{MAG\_MO\-DEL}}).
Moreover, it is also possible to measure morphological parameters of the galaxies, such as spheroid effective radius, disk aspect ratio, and disk-scale length. \\
The model of the PSF may be employed to extract a more accurate star/galaxy classification using the new SExtractor classifier, {\tt{SPREAD\_\-MODEL}}, which is a normalized, simplified linear discriminant between the best fitting local PSF model and a more extended model made by the same PSF convolved with a circular exponential disk model with scalelength = FWHM/16, where FWHM is the full-width at half maximum of the PSF model \citep{desai12}. \\
A more detailed description of PSFEx and the new SExtractor capabilities can be found in \citet{bertin11} and \citet{armstrong2010}.\\
Throughout the paper, we indicate simply with PSFEx the software version 3.9.1.

\section{Catalog extraction}
\label{4}

In this section we provide a general discussion on how to set the input parameters in order to extract the catalogs with the software tools presented above.\\

\subsection{DAOPHOT and ALLSTAR}
\label{4.1}

Besides instrumental parameters, such as gain, saturation level, and readout noise, which are set according to the values used for the simulations (Sect.~\ref{2}), in DAOPHOT the detection and photometric options must be configured by means of input setup files.
In the present analysis, the threshold value was chosen to detect as many possible sources, while avoiding as much as possible spurious detections. In fact, as the threshold decreases, the number of detected sources increases up to a certain value, for which the relation changes in steepness. Thus it is possible to choose a reasonable value for the threshold, by plotting the number of extracted sources for different threshold values and by choosing the threshold near the ``elbow'' of the function. Moreover, in order to avoid spurious detections, due to Poissonian noise, the extracted catalog was visually inspected.  In fact, by using the input parameters as reported in Tab~\ref{tab1}, only the 5\% of the objects extracted in the catalog is spurious. By decreasing the threshold, although the number of extracted input sources increases, the percentage of spurious detections rapidly grows up to 30\% at 4$\sigma$ threshold, even reaching more then 88\% at 3$\sigma$. \\
The {\tt{FITTING RADIUS}} was set equal to the FWHM of the image (see Tab.~\ref{tab1}). \\
To obtain the best aperture radius, we have derived the growth curve for input stellar sources. Then, we fixed the aperture radius to 12.5 pixel (see \texttt{a1} in Tab.~\ref{tab1}), producing the better coverage of the sources input magnitude.
Thus, values of {\tt{INNER RADIUS}} and {\tt{OUTER RADIUS}} were chosen accordingly, smaller and greater than the aperture radius, respectively.\\
The PSF analytical model was chosen with the higher level of complexity, that is, the implementations of the Penny function with five free parameters (option 6). We chose to visually inspect the image of the PSF produced by DAOPHOT for all the PSF stars. ALLSTAR parameters were set accordingly to those established for DAOPHOT and, furthermore, we required the redetermination of the centroids. \\
The main parameters set for DAOPHOT and ALLSTAR are reported in Tab.~\ref{tab1}.

\begin{table}[h]
\centering
\small
 \begin{tabular}{ l r }
 \hline
 \hline
\rule[-1.0ex]{0pt}{2.5ex}  Parameter & Values\\
\hline
{\tt{FITTING RADIUS}} & 3.38\\
{\tt{THRESHOLD (in sigmas)}} & 5\\
{\tt{ANALYTIC MODEL PSF}} & 6\\
{\tt{PSF RADIUS}} & 7.5\\
{\tt{a1}} & 12.5\\
{\tt{INNER RADIUS}} & 20\\
{\tt{OUTER RADIUS}} & 35\\
{\tt{REDETERMINE CENTROIDS}} & 1.00\\
\hline
\end{tabular}
\caption{Main input parameters set in the DAOPHOT and ALLSTAR configuration files. {\tt{FITTING RADIUS}}, {\tt{PSF RADIUS}}, {\tt{a1}}, {\tt{INNER RADIUS}} and {\tt{OUTER RADIUS}} are expressed in pixels.}
\label{tab1}
\end{table}

\subsection{SExtractor and PSFEx}
\label{4.2}

As for DAOPHOT, SExtractor instrumental parameters are set accordingly to those defined as input in the simulations (see Sect.~\ref{2}). \\
Concerning the sky background modeling and subtraction, we decided to automatically estimate the background within the software package adopting the global background map. Given the average size of the objects, in pixels, in our images, we chose to leave {\tt{BACK\_SIZE}} to the default value 64.
The choice of the filter was more complex. We performed several tests with various filters obtaining the best results by Gaussian and top-hat masks.
However, the choice between the various filters, although affecting the number of detected sources, does not alter their measurements.\\
For the thresholding parameters, we followed the same procedure approached with DAOPHOT, as described in Sect.~\ref{4.1}, that is, by choosing a value near to the change in gradient of the relation between the number of extracted sources and the threshold value for detections. Moreover, the catalog was visually inspected to avoid residual spurious detections and to verify the deblending parameters.\\
We fixed the size of the aperture for photometry, according to the one set in DAOPHOT, to 25 pixels of diameter (\texttt{PHOT\_APERTURES}). For PSFEx parameters we used a set of 20 pixel basis and a size for the PSF image of 25 pixels according with the aperture size. We adopted a 25$\mathrm{\times}$25 pixel kernel following PSF variations within the image up to $\mathrm{2^{nd}}$ order. The main values set for SExtractor and PSFEx are reported in Tab.~\ref{tab2}.\\

\begin{table}[h]
\centering
\small
\begin{tabular}{ l r }
\hline
\hline
\rule[-1.0ex]{0pt}{2.5ex}  Parameter & Values\\
\hline
{\tt{DETECT\_MINAREA}} & 5\\
{\tt{DETECT\_THRESH}} & 1.5$\sigma$ \\
{\tt{ANALYSIS\_THRESH}} & 1.5$\sigma$ \\
{\tt{FILTER\_NAME}} & tophat\_3.0\_3x3.conv\\
{\tt{DEBLEND\_NTHRESH}} & 64\\
{\tt{DEBLEND\_MINCON}}T & 0.001\\
{\tt{BACK\_TYPE}} & GLOBAL\\
{\tt{BACK\_SIZE}} & 64\\
{\tt{BACK\_FILTERSIZE}} & 3\\
{\tt{PHOT\_APERTURES}} & 25\\
{\tt{BASIS\_TYPE}} & PIXEL\_AUTO\\
{\tt{BASIS\_NUMBER}} & 20\\
{\tt{PSF\_SIZE}} & 25,25\\
\hline
\end{tabular}
\caption{Main input parameters set in the SExtractor and PSFEx configuration files. {\tt{DETECT\_MINAREA}}, {\tt{BACK\_SIZE}}, {\tt{PHOT\_APERTURES}} and {\tt{PSF\_SIZE}} are expressed in pixels.}
\label{tab2}
\end{table}

\section{Results}
\label{5}

\begin{figure}[!ht]
\centering
\includegraphics[scale=0.4]{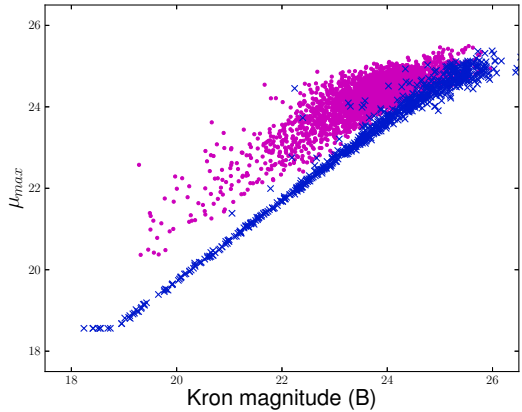}
\caption{$\mathrm{\mu_{max}}$ as a function of the Kron magnitude for stars (diagonal crosses) and galaxies (points) in the SExtractor catalog.}
\label{fig1}
\end{figure}

In this section, we compare the results obtained using the two software packages. We will focus on four aspects of the extracted catalog namely: photometric depth, reliability, accuracy of the derived photometry, and determination of the positions of the centroids.\\
All the quantities and the statistics shown in this section are obtained by excluding saturated sources. In Fig.~\ref{fig1}, it is shown $\mathrm{\mu_{max}}$ as a function of the Kron magnitude of the objects extracted by SExtractor. As evidenced by the flattening of star sequence, sources with magnitude B$\mathrm{\le}$19 mag are saturated in the simulated images. Starting from Tab.~\ref{tab3} and Fig.~\ref{fig2}, we report the comparison among results obtained for the whole input magnitude range of unsaturated sources: 19-26 mag. However, since we consider only input stellar sources recovered by both software packages and since the completeness limit of the DAOPHOT star catalog is B=24 mag (see Sect.~\ref{5.1}), the last two reported magnitude bins are underpopulated and the results may be affected by catalog incompleteness.

\subsection{Photometric depth}
\label{5.1}

\begin{figure}[!ht]
\centering
\includegraphics[scale=0.4, clip=true]{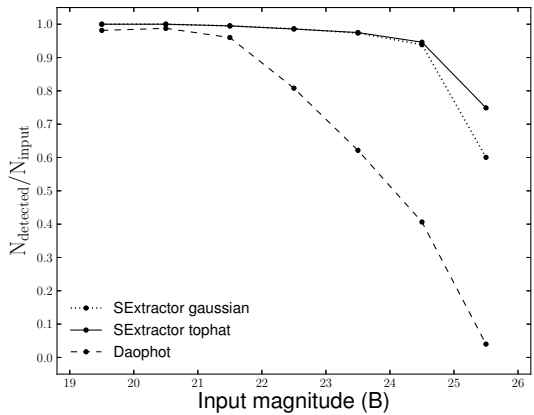}
\caption{Ratio between detected and input sources for different magnitude the bins. The dotted and the solid lines refer to SExtractor used with a Gaussian and a top-hat filter respectively, while the dashed line refers to values obtained with DAOPHOT.}
\label{fig2}
\end{figure}

\begin{figure*}[!ht]
\centering
\subfloat[]{
\label{fig3a}
\includegraphics[scale=0.4]{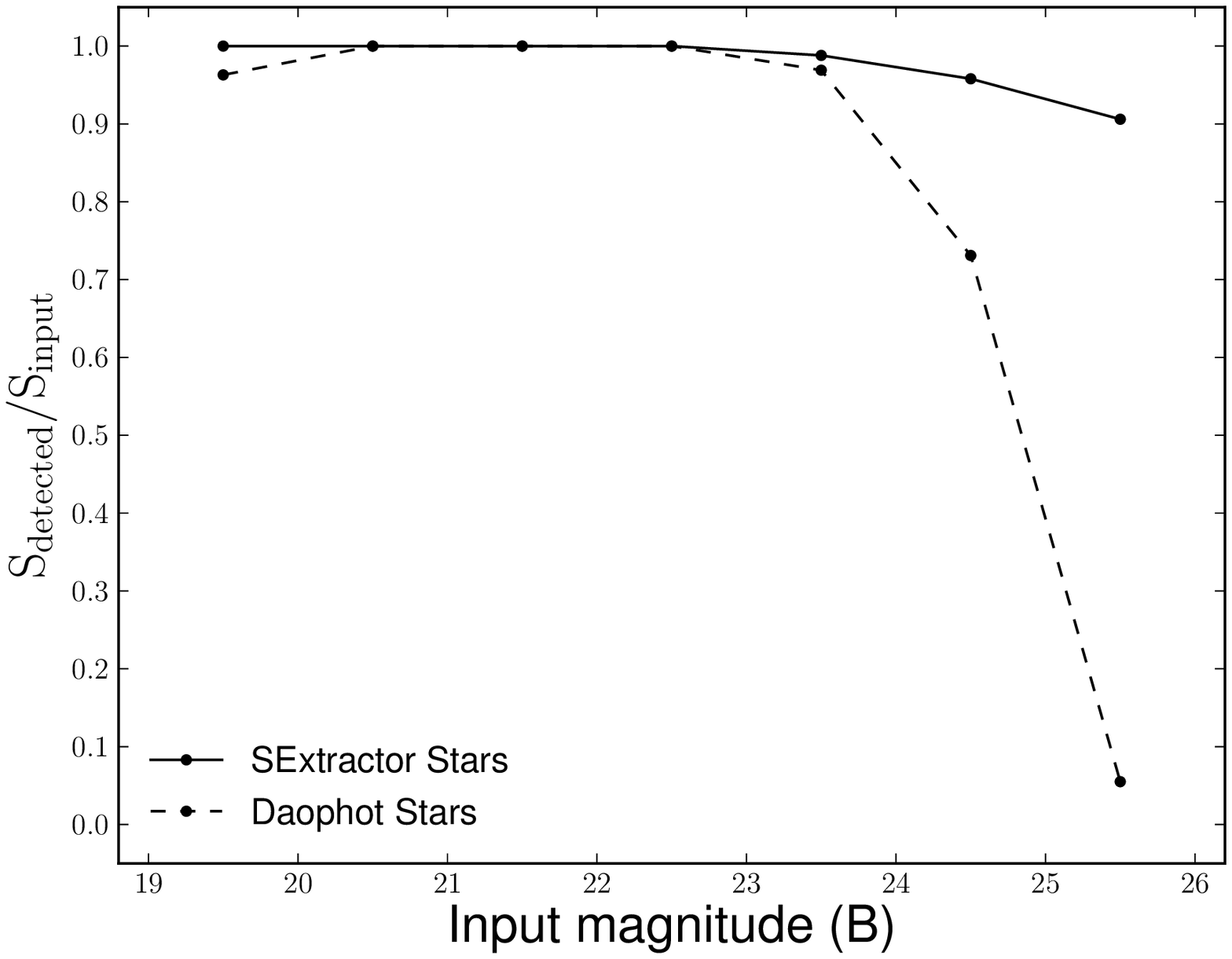}}
\subfloat[]{
\label{fig3b}
\includegraphics[scale=0.4]{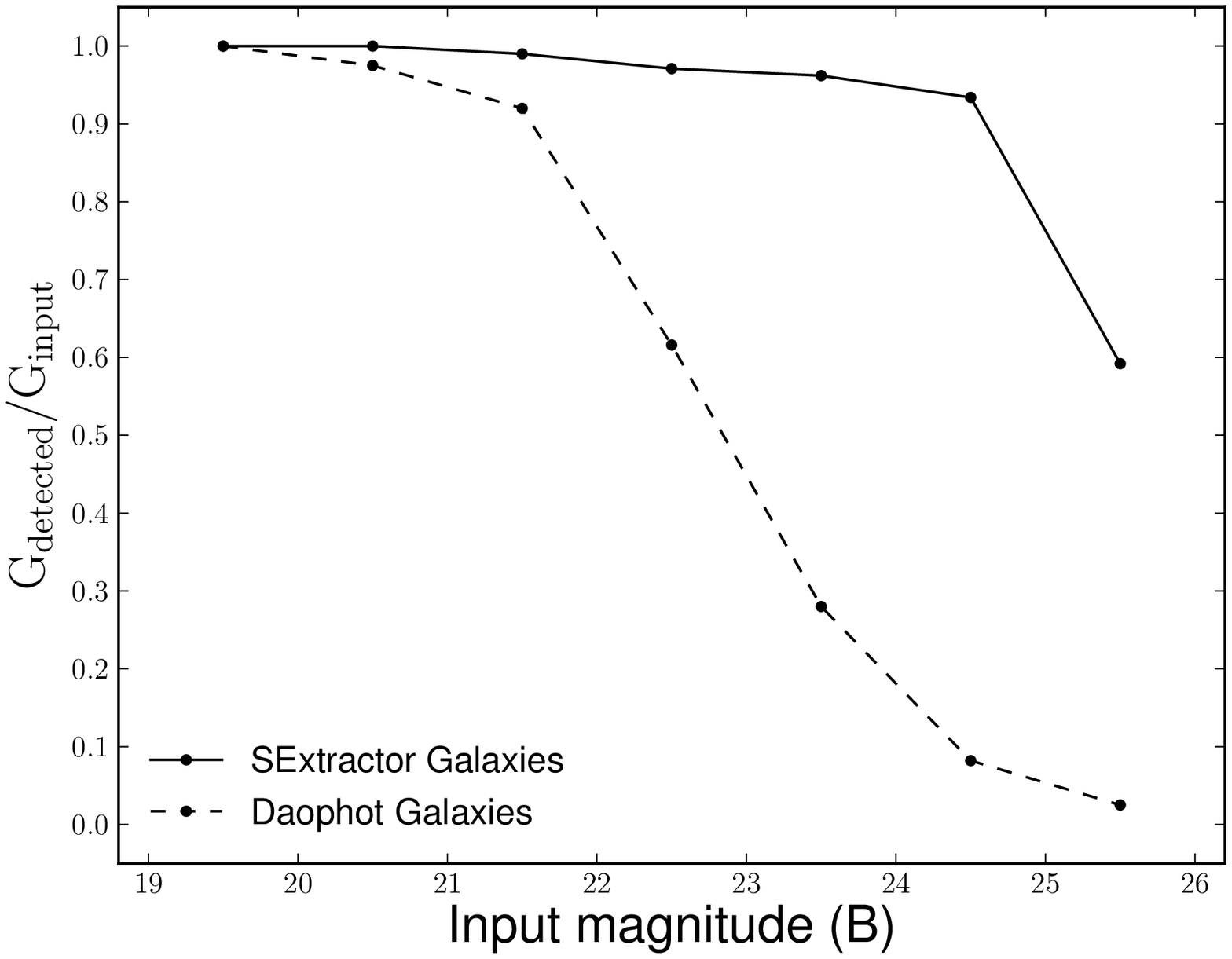}}
\caption{The left panel shows the ratio between the numbers of detected and input stars as a function of magnitude bins, as obtained by SExtractor (solid line) and by DAOPHOT (dashed line); in the right panel are plotted the same quantities, but for galaxies.}
\end{figure*}

The photometric limiting magnitude of the extracted catalog is defined as the magnitude limit below which the completeness drops down to 90\%, where the completeness is the ratio of number of detected sources, $\mathrm{N_{detected}}$, and number of input sources, $\mathrm{N_{input}}$.\\
With DAOPHOT, the photometric depth depends mainly on the threshold applied, while for SExtractor, it depends also on the deblending of the sources, and on the filter used for the detection (see Sect.~\ref{3}). As discussed in Sect~\ref{4.1} and~\ref{4.2}, in order to fix the thresholding and deblending parameters, we performed several tests, by visually inspecting the extracted sources, and finally we fixed the values reported in Tab.~\ref{tab1} and~\ref{tab2}. Then, we compared the results of source extraction obtained using two different filters: a Gaussian (dotted line in Fig.~\ref{fig2}) and a ``top-hat'' function (continuous line in Fig.~\ref{fig2}). As shown in Fig.~\ref{fig2}, using SExtractor with a top-hat filter, we can improve the detection of faint sources. In this case, the depth of the catalog is $\sim$ 25.0 mag. Hence, we refer to this filter in all the tests performed with SExtractor and reported below. Figure~\ref{fig2} shows also the percentage of extracted sources per magnitude bin obtained using DAOPHOT (dashed line). With this software package, the completeness drops rapidly to very low values for magnitudes fainter than B = 22.0 mag. However, this comparison is misleading. In fact, DAOPHOT is not designed to work with extended sources.
For this reason, in Fig.~\ref{fig3a} we report the ratio between the detected sources, which are \textit{a priori} known to be stars ($\mathrm{S_{detected}}$), and the input stars ($\mathrm{S_{input}}$). In Fig.~\ref{fig3b} we show the same quantities but for galaxies ($\mathrm{G_{detected}}$, $\mathrm{G_{input}}$).
We can see that the fraction of detected source is higher for stars for both SExtractor (B = 26.0 mag) and DAOPHOT (B = 24.0 mag). Hence, in conclusion, considering only stars, the final depth returned by DAOPHOT is $\sim$ 2 mag brighter than those produced by SExtractor. \\

\subsection{Reliability of the catalog}
\label{5.2}

\begin{figure*}
\centering
\subfloat[]{
\centering
\label{fig4a}
\includegraphics[scale=0.4]{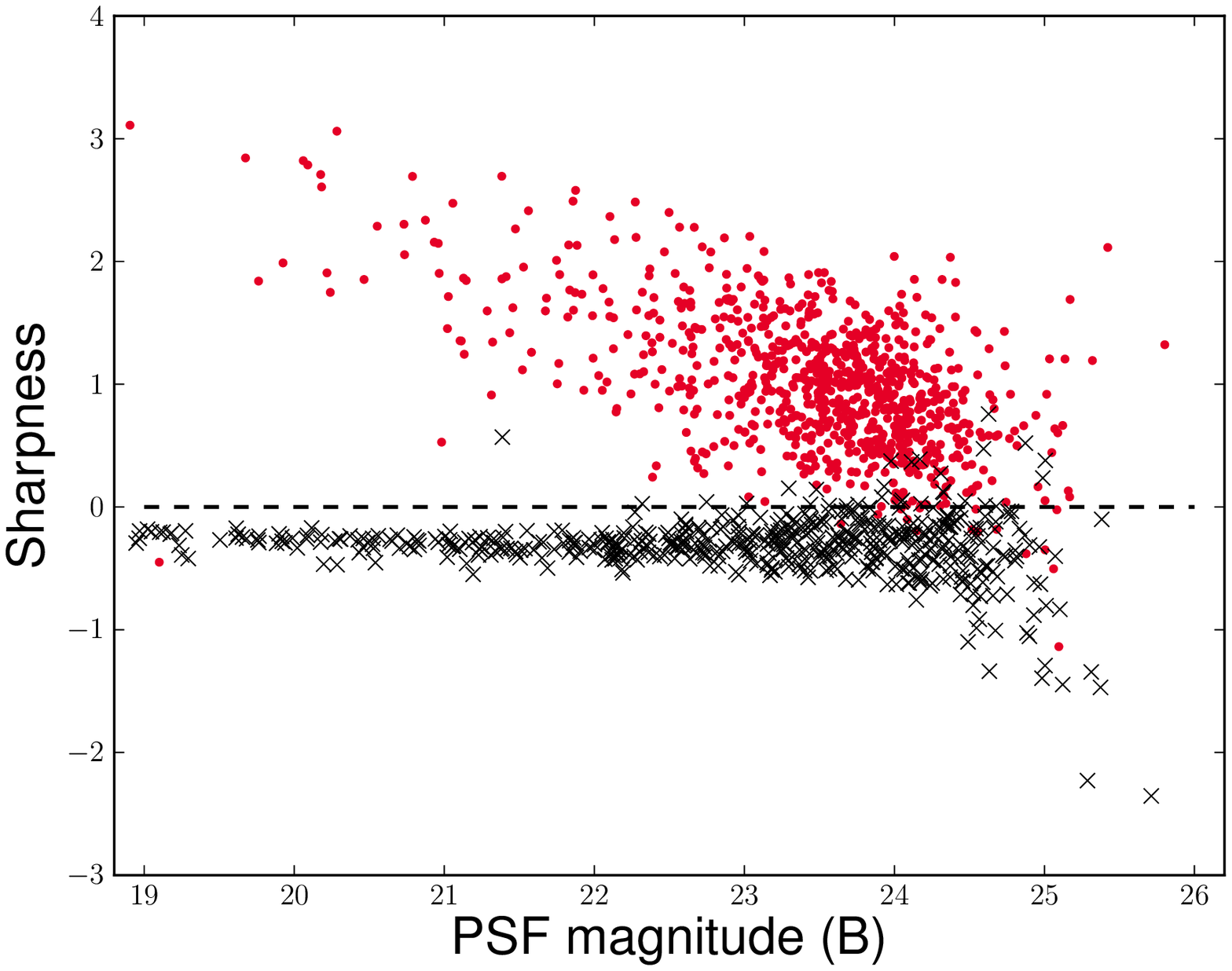}}
\subfloat[]{
\centering
\label{fig4b}
\includegraphics[scale=0.4]{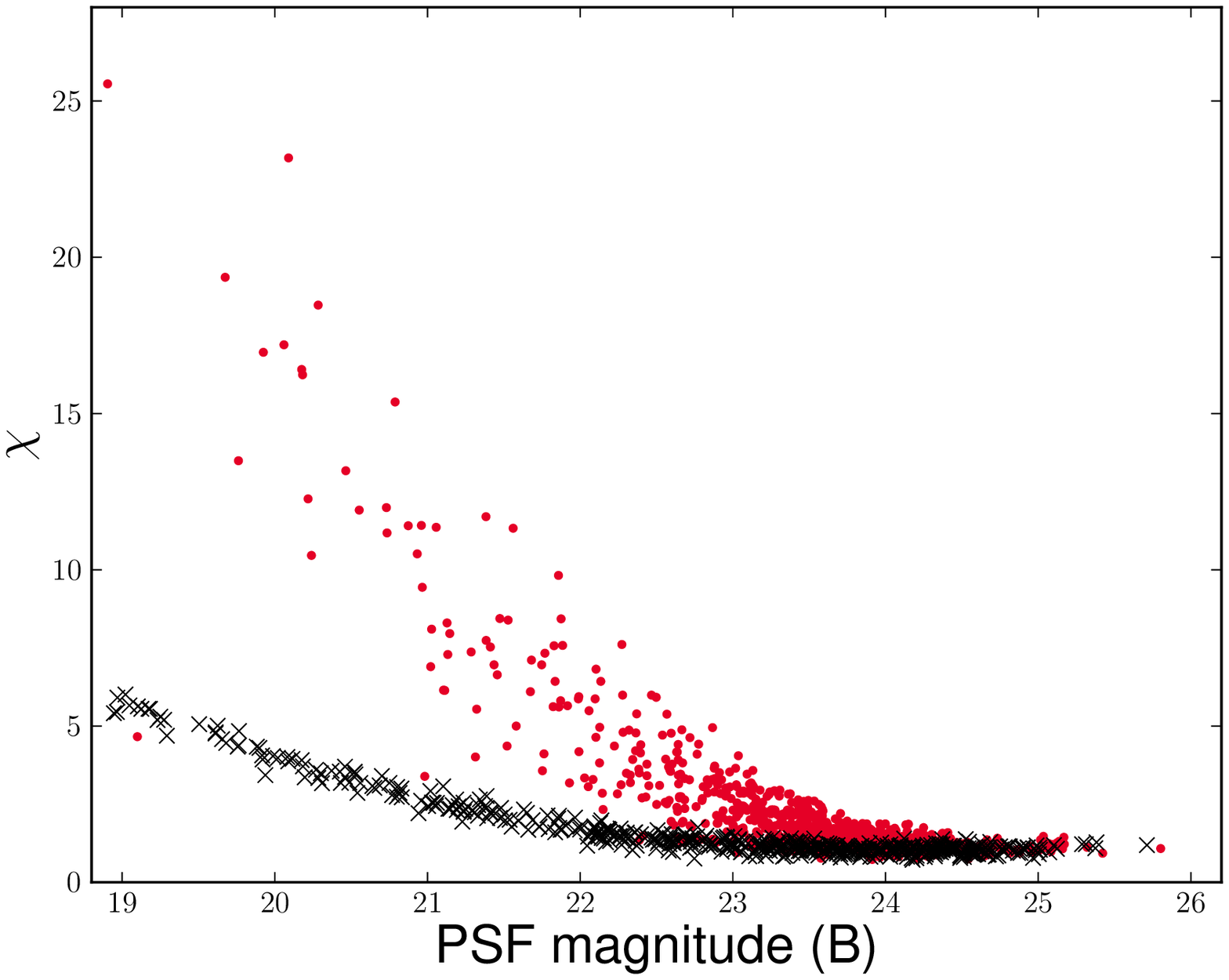}}
\label{fig4}
\caption{Distribution of DAOPHOT sharpness ({\it left panel}) and $\chi$ ({\it right panel}) as a function of the PSF magnitude for simulated stars (diagonal crosses) and galaxies (points). In the left panel the dashed line at zero sharpness is the adopted separation limit for the star/galaxy classification (see Sect.~\ref{5.2}).}
\end{figure*}

\begin{figure*}
\centering
\subfloat[]{
\includegraphics[scale=0.38]{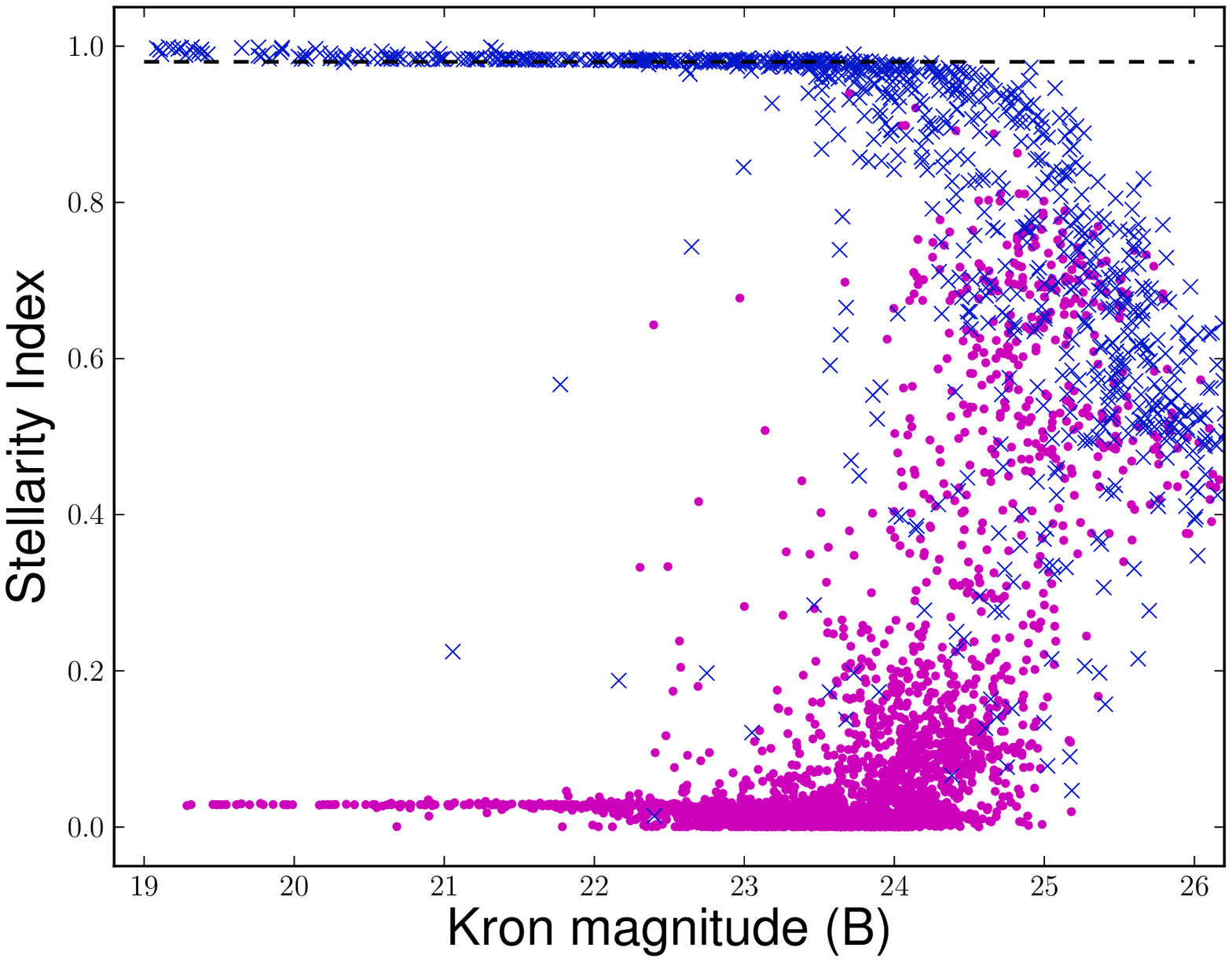}
\label{fig5a}}
\subfloat[]{
\includegraphics[scale=0.38]{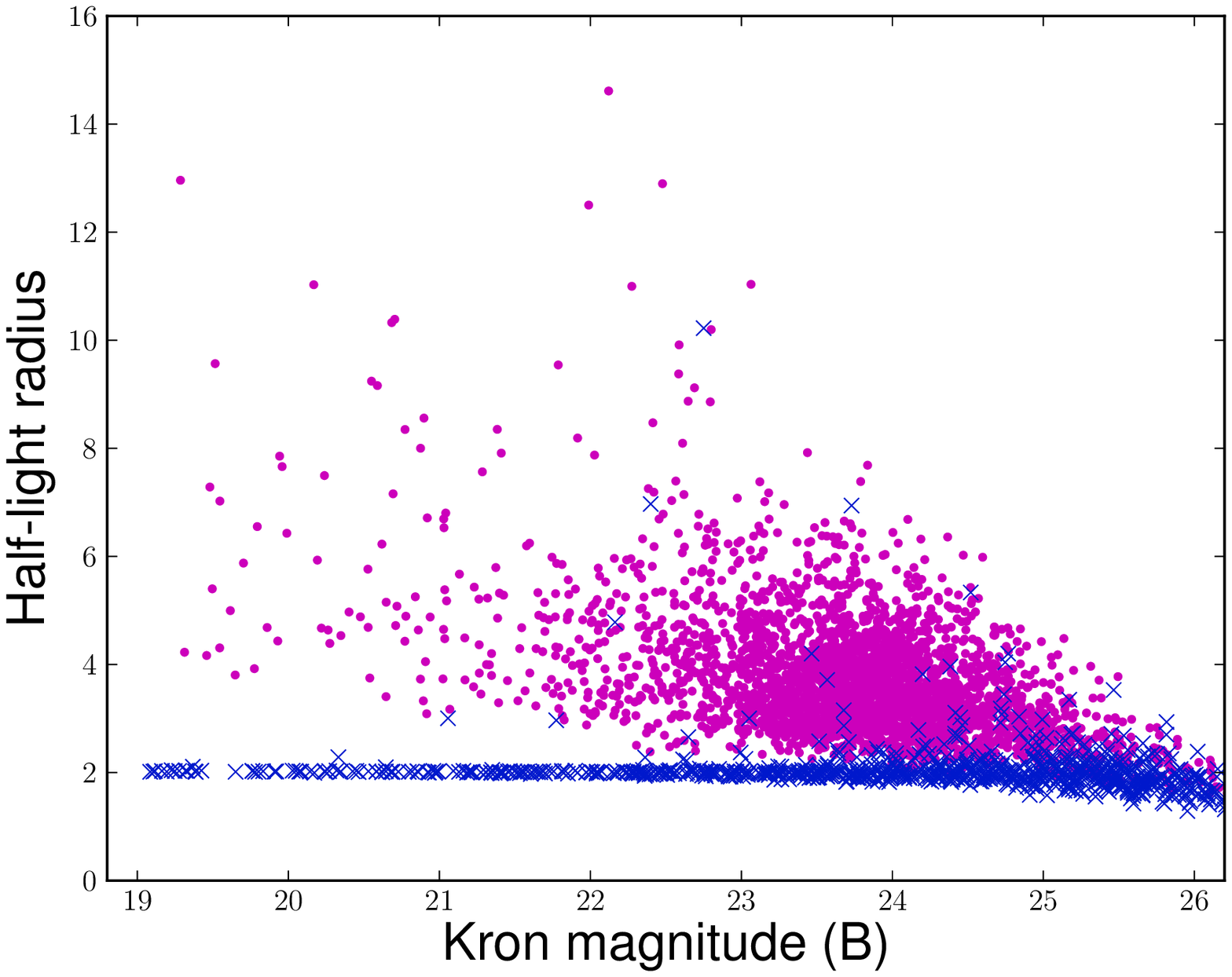}
\label{fig5b}}
\\
\subfloat[]{
\includegraphics[scale=0.38]{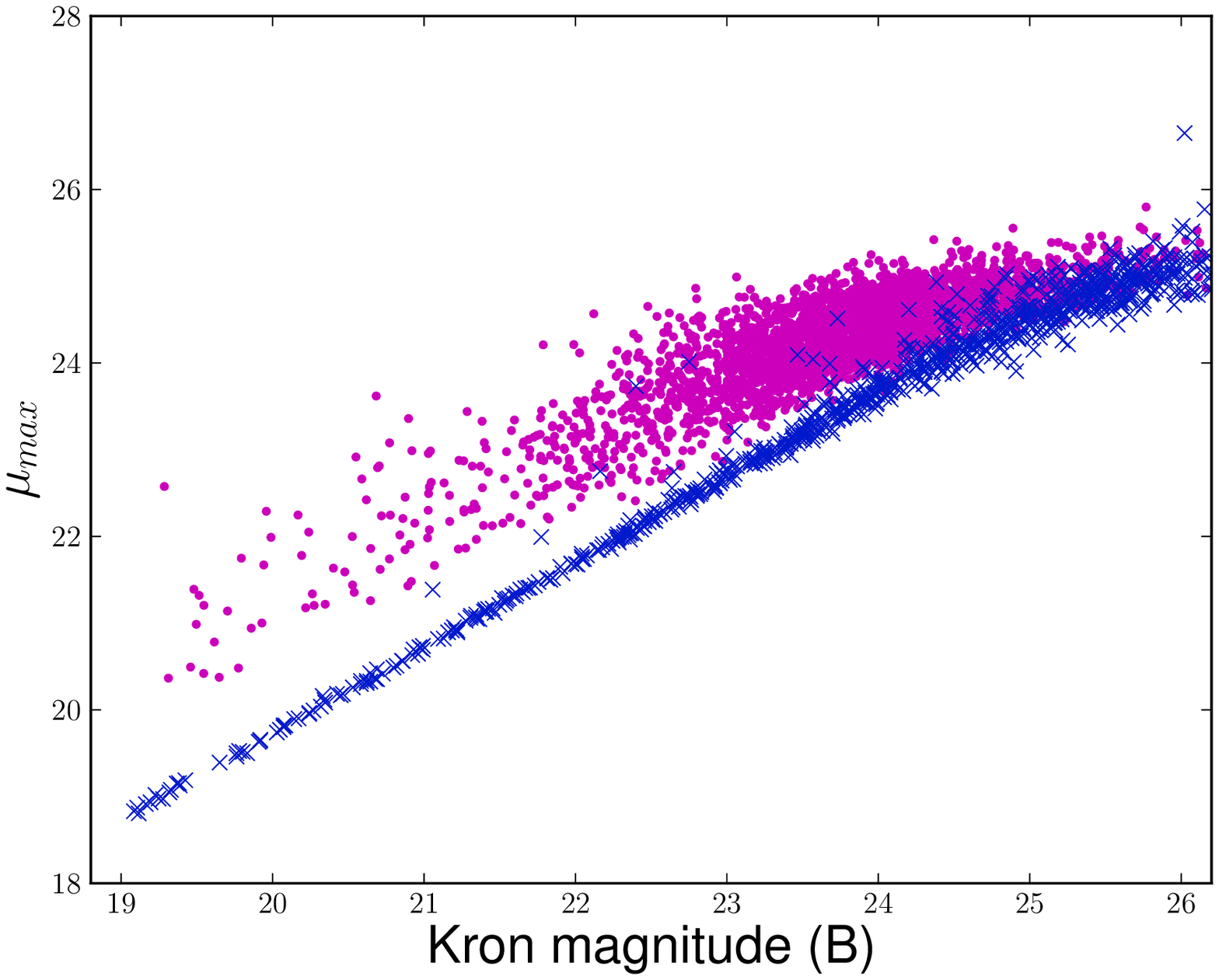}
\label{fig5c}}
\subfloat[]{
\includegraphics[scale=0.38]{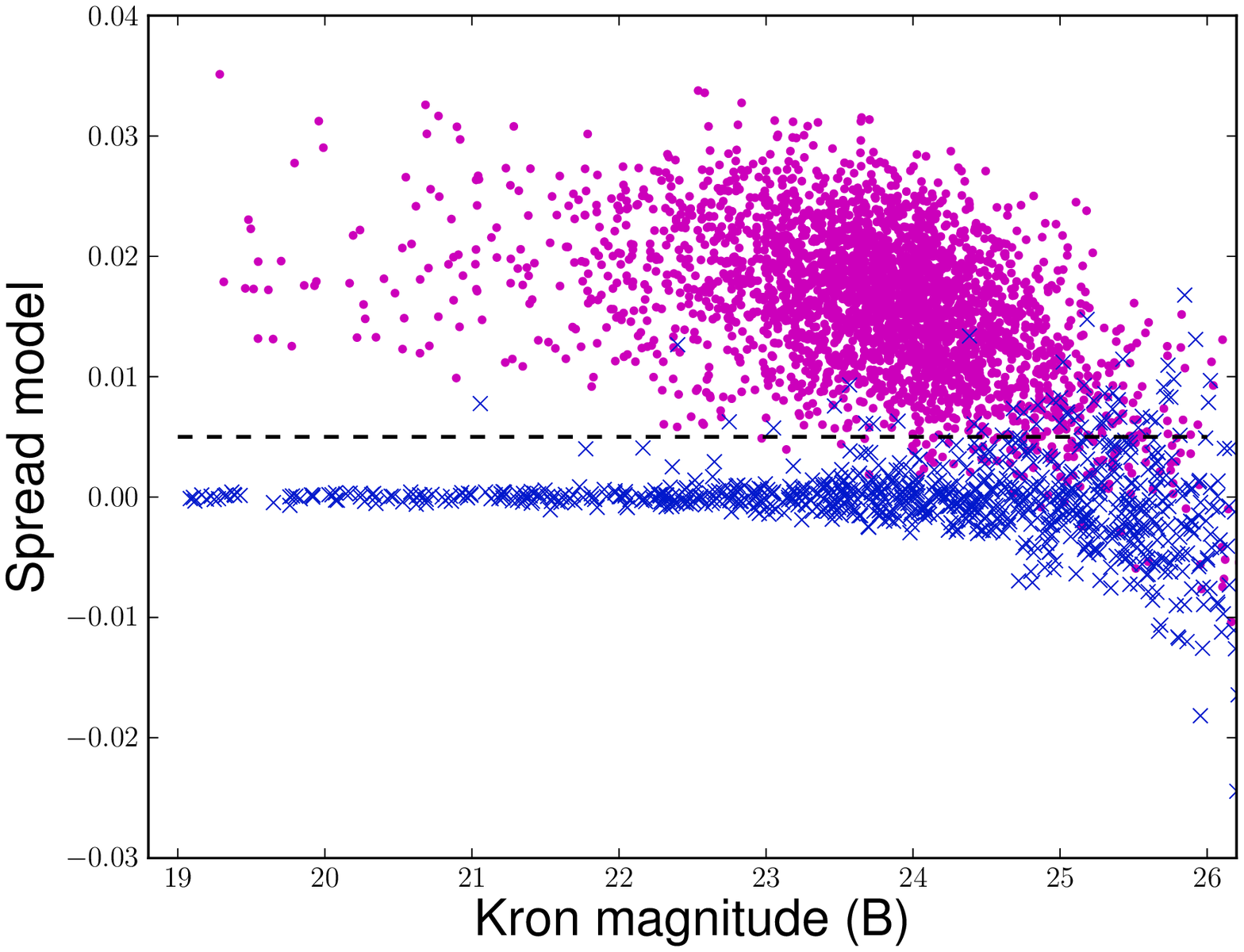}
\label{fig5d}}
\caption{Distribution of SExtractor stellarity index ({\it{panel a}}), half-light radius ({\it{panel b}}), $\mathrm{\mu_{max}}$ ({\it{panel c}}) and spread model ({\it{panel d}}), as a function of the Kron magnitudes for simulated stars (diagonal crosses) and galaxies (points). The dashed line in {\it{panels a}} and \textit{d} is the adopted separation limit for the star/galaxy classification (see Sect.~\ref{5.2}). }
\end{figure*}

The reliability of the catalog is defined as the ratio between the number of well-classified sources and the number of the sources detected by the software packages (see Eq. (6) of \citealt{laher08}).
For these tests we use only the set of stars detected ($\mathrm{S_{detected}}$) and well-classified ($\mathrm{S_{classified}}$) by both SExtractor and DAOPHOT. We compared results obtained with several methods to classify the sources. In fact, each method leads to a different estimate of the reliability. \\
As far as DAOPHOT is concerned, we used the output parameters \texttt{SHARP} (see Fig.~\ref{fig4a}) and $\chi$ (see Fig.~\ref{fig4b}), made the determination with ALLSTAR (see Sect.~\ref{3.1}). Fig.~\ref{fig4a} shows the distribution of ALLSTAR sharpness {\tt{SHARP}} for our data. The separation between the two classes seems to be well defined. On the other hand, Fig.~\ref{fig4b} shows that the use of the $\chi$ parameter does not improve the star/galaxy classification. For this reason, we classified as stars all the sources with {\tt{SHARP}} lower than 0.\\
In order to investigate the reliability of the catalog made with SExtractor, we used both traditional methods as well as the new parameter {\tt{SPREAD\_MODEL}}. In Fig.~\ref{fig5a}, we plot {\tt{CLASS\_STAR}} as a function of the Kron magnitude for our data. As shown, the lower the established limit to separate stars and galaxies, the higher will be the contamination of the star subsample from galaxies. A reasonable limit for the separation is 0.98.\\

\begin{figure}[!ht]
\centering
\includegraphics[scale=0.4]{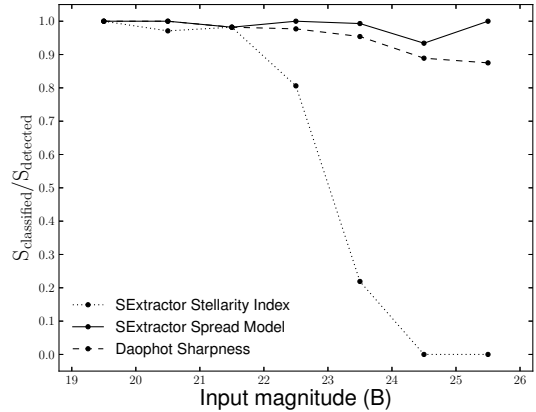}
\caption{Ratio between stars classified by Stellarity Index (dotted line) and Spread Model (solid line) from SExtractor with threshold values respectively to 0.98 and 0.005, and by DAOPHOT sharpness (dashed line) with a threshold value equal to zero and detected stars, as function of input magnitude.}
\label{fig6}
\end{figure}

Figures~\ref{fig5b} and~\ref{fig5c} show the locus of stars, selected according to the relation between half-light radius and $\mu_{max}$, respectively, as a function of the Kron magnitude. There is an improvement of the source classification compared to the use of {\tt{CLASS\_STAR}} parameter, allowing a reliable star/galaxy separation down to B $\mathrm{=}$ 23.5 mag.\\
Finally, Fig.~\ref{fig5d} shows {\tt{SPREAD\_MODEL}} values as a function of Kron magnitude.
Stars and galaxies tend to arrange themselves in two distinct places on the plot. Also in this case, the higher we choose the separation limit, the higher will be the contamination of the stellar sequence from galaxies. A good compromise between a reliable classification and a low contamination is the value 0.005. \\
In Fig.~\ref{fig6}, it is shown the ratio between the sources correctly classified as stars using the stellarity index (dotted line), spread model (continuous line) and sharpness parameter (dashed line), as function of input magnitude. \\
In conclusion, if we define a classification with a reliability of at least 90\% with these methods we can acceptably classify the stars in DAOPHOT down to about 24 mag, which is the photometric depth of the extracted catalog, while in the case of SExtractor, the classifier {\tt{SPREAD\_MODEL}} allows us to obtain a reliable star/galaxy separation down to B $\mathrm{=}$ 26 mag.\\

\subsection{Photometry}
\label{5.3}

\begin{figure*}[!ht]
 \centering
 \subfloat[]{
 \includegraphics[scale=0.4]{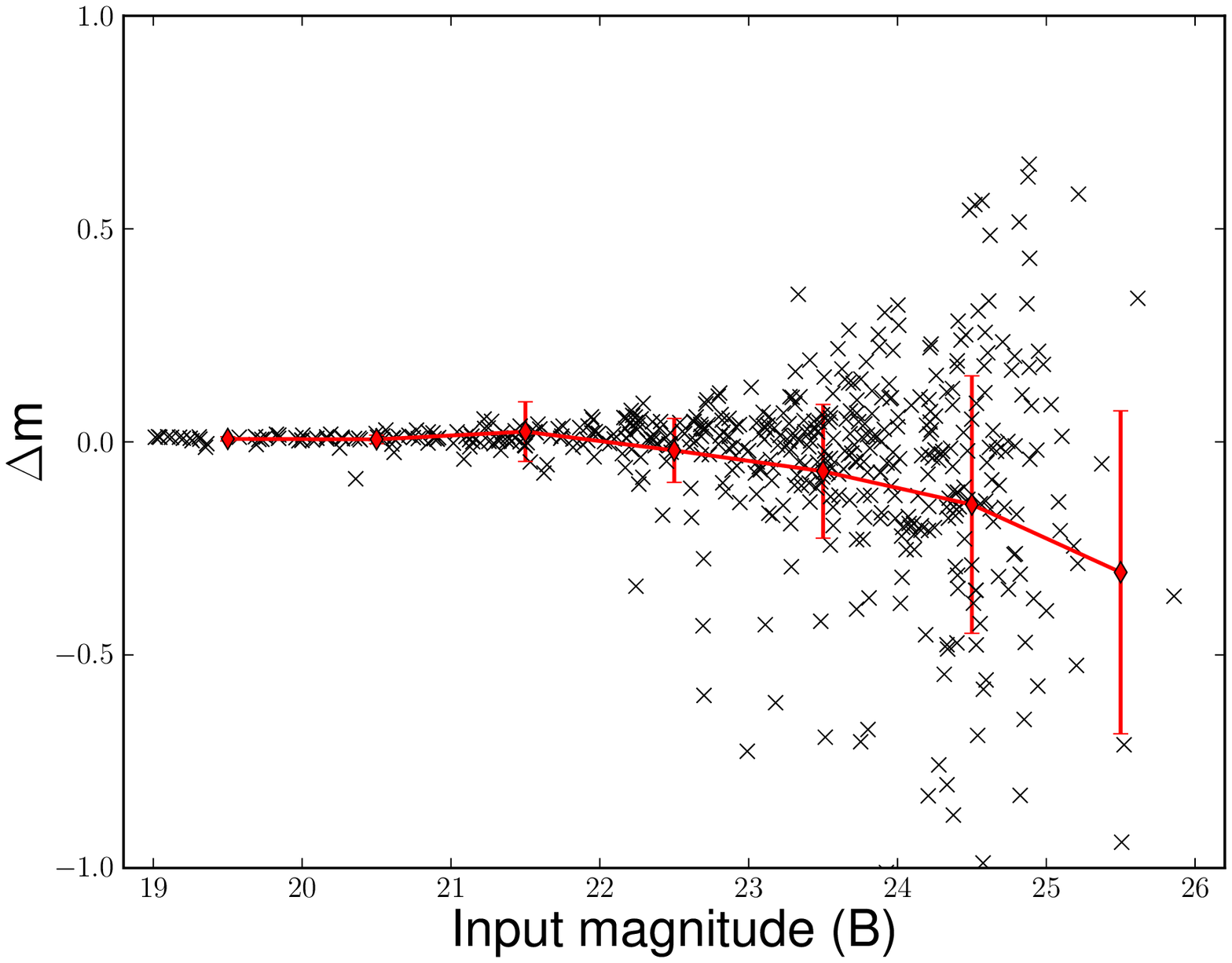}
 \label{fig7a}
 }
 \subfloat[]{
 \includegraphics[scale=0.4]{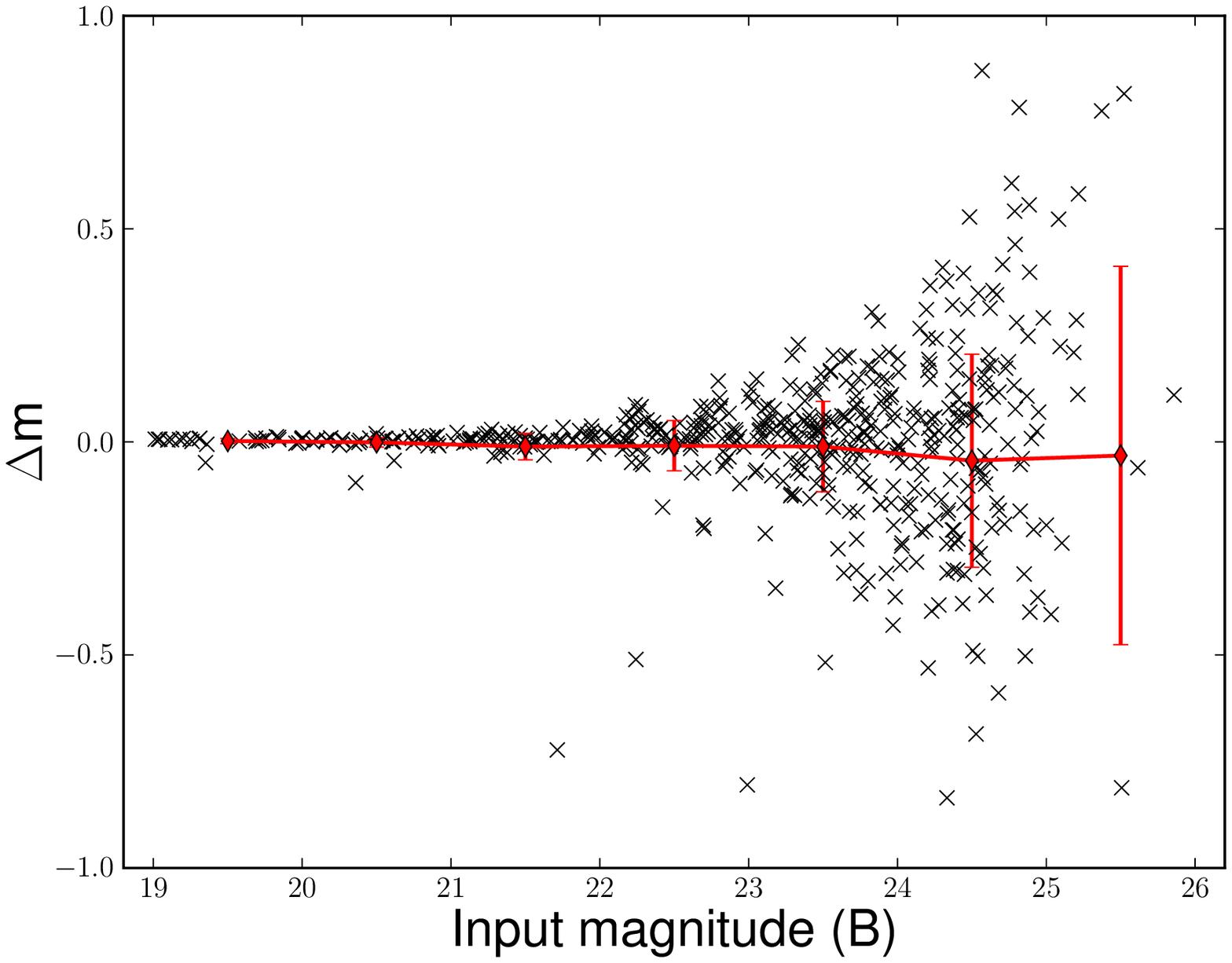}
 \label{fig7b}
 }
 \\
 \subfloat[]{
 \includegraphics[scale=0.4]{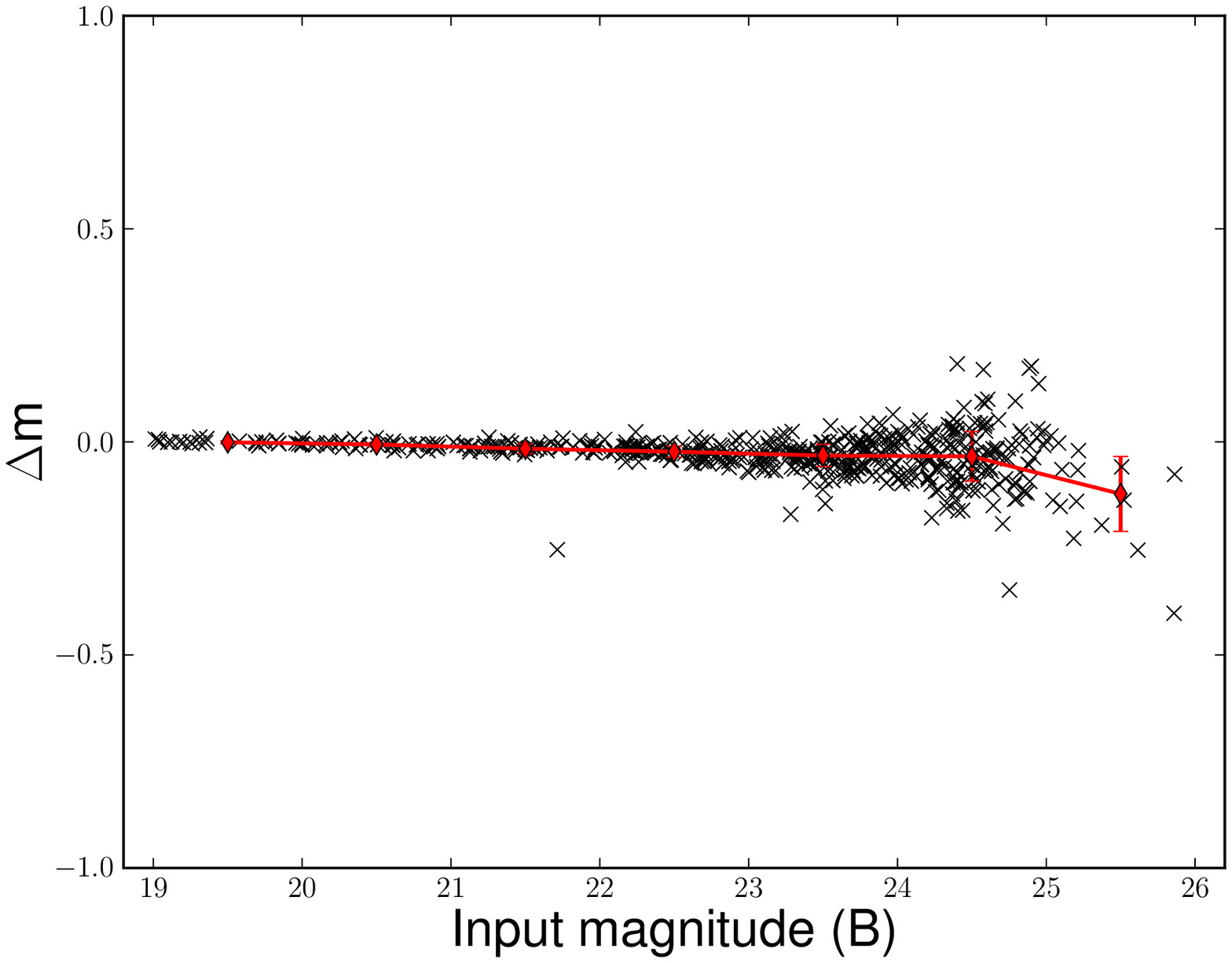}
 \label{fig7c}
 }
 \subfloat[]{
 \includegraphics[scale=0.4]{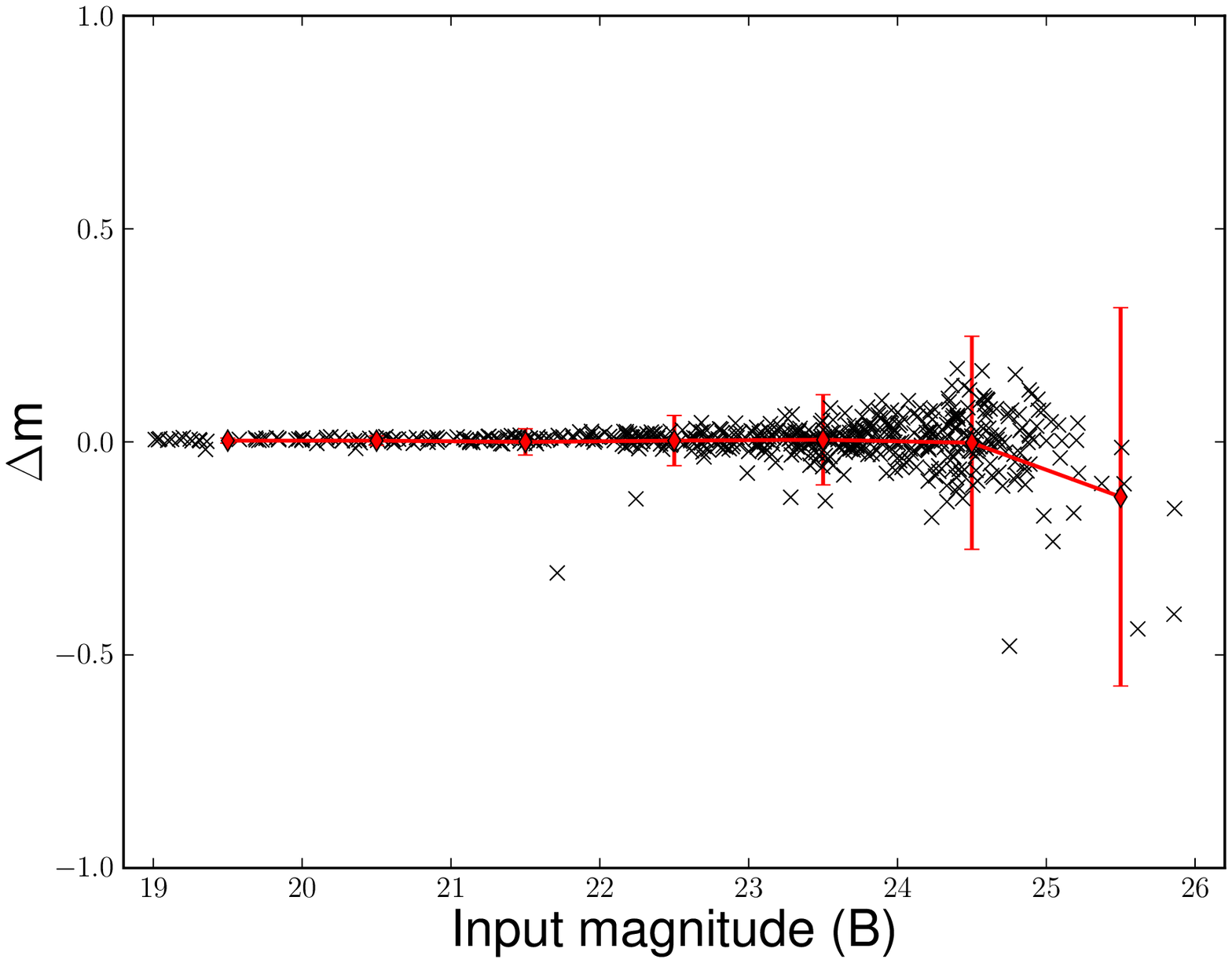}
 \label{fig7d}
 }
 \caption{\textit{Top panels}: Residuals between aperture magnitudes estimated by DAOPHOT ({\it{left panel}}) and by SExtractor ({\it{right panel}}), and input magnitudes for detected stars. \textit{Bottom panels}: Residuals between PSF magnitude estimated by DAOPHOT ({\it{left panel}}) and by SExtractor ({\it{right panel}}), and input magnitude for detected stars. Superimposed red points and solid red lines draw the mean and standard deviation values reported in Tab.~\ref{tab3}.}
 \label{fig7}
 \end{figure*}

In this section, we compare the results obtained with aperture and PSF photometry on the sample of stars detected by both SExtractor and DAOPHOT. We also investigate the results obtained with Kron, isophotal and model-fitting photometry for galaxies detected by SExtractor. \\
In Tab.~\ref{tab3}, we report the mean difference and the standard deviation between aperture and PSF magnitudes, as estimated by DAOPHOT (parts a and b, respectively) and SExtractor (parts c and d, respectively), against input magnitude.\\
Figure~\ref{fig7} shows the residuals between aperture and input magnitudes (top panels), and the residuals between PSF and input magnitudes (bottom panels), as estimated by DAOPHOT (left panels) and SExtractor (right panels).\\
Table~\ref{tab3} and Fig.~\ref{fig7} show that there is a characteristic broadening of the residuals at fainter magnitudes, as we expect when measurements become sky-noise dominated, but the spread in the case of PSF photometry remains smaller than for aperture measurements.
This behavior is well known (e.g. \citealp{becker07}) for DAOPHOT, but it is worth underlining that SExtractor has reached this level of accuracy in PSF photometry only after the release of PSFEx.\\
In the top part of the Tab.~\ref{tab3}, we also report the mean difference and the standard deviation between Kron (part a), isophotal (part b) and model magnitudes (part c), respectively, and input magnitudes for stars.\\
For completeness, since SExtractor is designed also to obtain accurate galaxy photometry, we report in the bottom part of the Tab.~\ref{tab4} the mean difference and the standard deviation between Kron (part d), isophotal (part e) and model magnitudes (part f), and input magnitudes for the ``true'' galaxies detected by the software package (See Sect.~5.2).\\
By considering only stellar photometry, both software packages are able to deliver acceptable performances for both aperture and PSF photometry, up to a threshold two magnitudes brighter than the limiting magnitudes of the input simulated images, which is the completeness limit of the DAOPHOT catalog. Furthermore, the Kron magnitude yields $\sim$ 94\% of the total source flux within the adaptive aperture (\citealp{bertin96}) so, accordingly, we see a shift of $\sim$ 0.07 mag even in the brightest magnitude bin. On the other hand, the isophotal magnitude depends on the detection threshold and the model magnitudes (obtained through a sum of bulge plus disk), and produce an unbiased estimate of the total magnitude also for stars.\\
In conclusion, the new PSF modeling of SExtractor produces photometric measurements as accurate as those obtained with DAOPHOT.\\

 \subsection{Centroids}
 \label{5.4}

\begin{figure*}[!ht]
 \centering
 \subfloat[]{
 \includegraphics[scale=0.4]{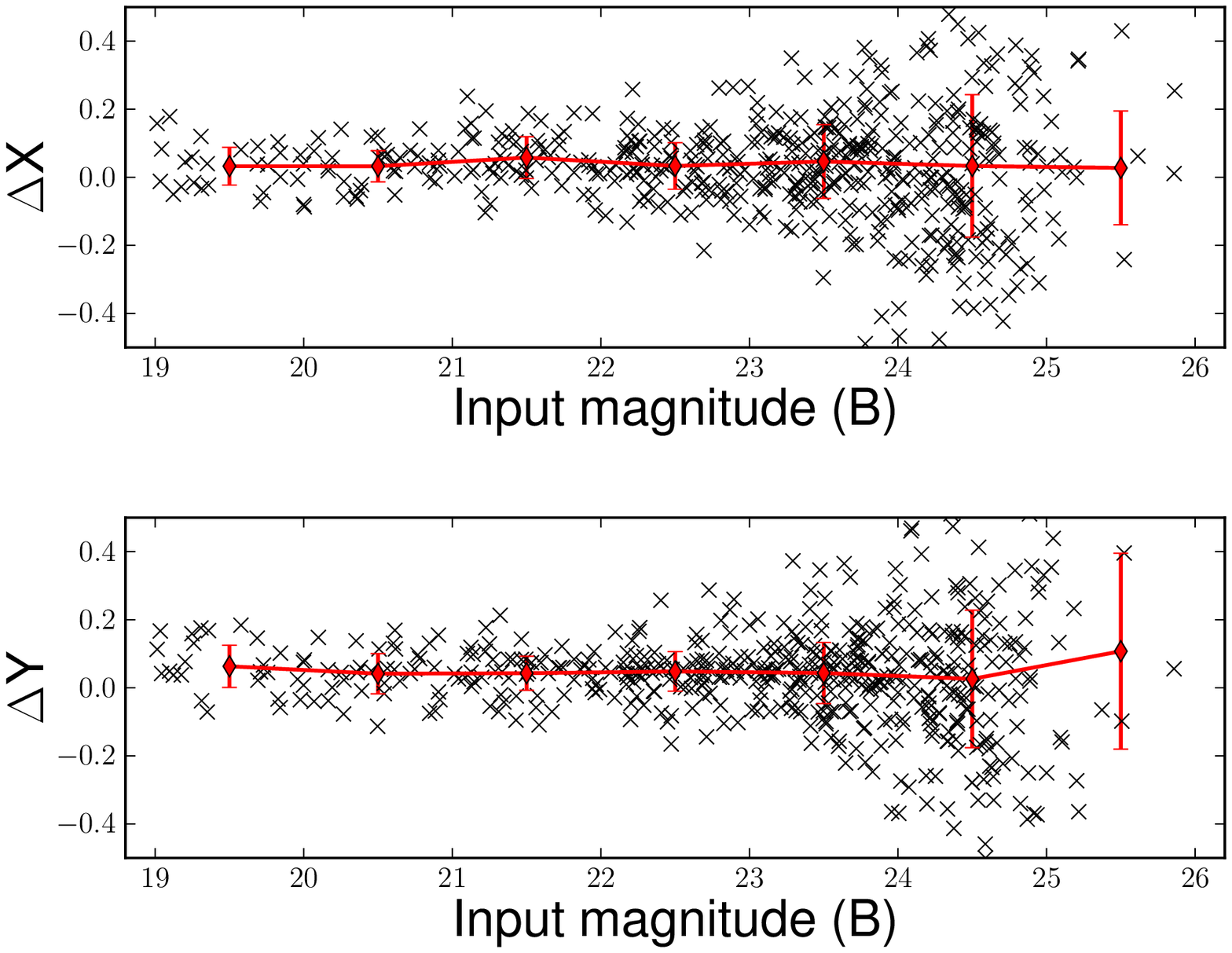}
 \label{fig8a}}
 \subfloat[]{
 \includegraphics[scale=0.4]{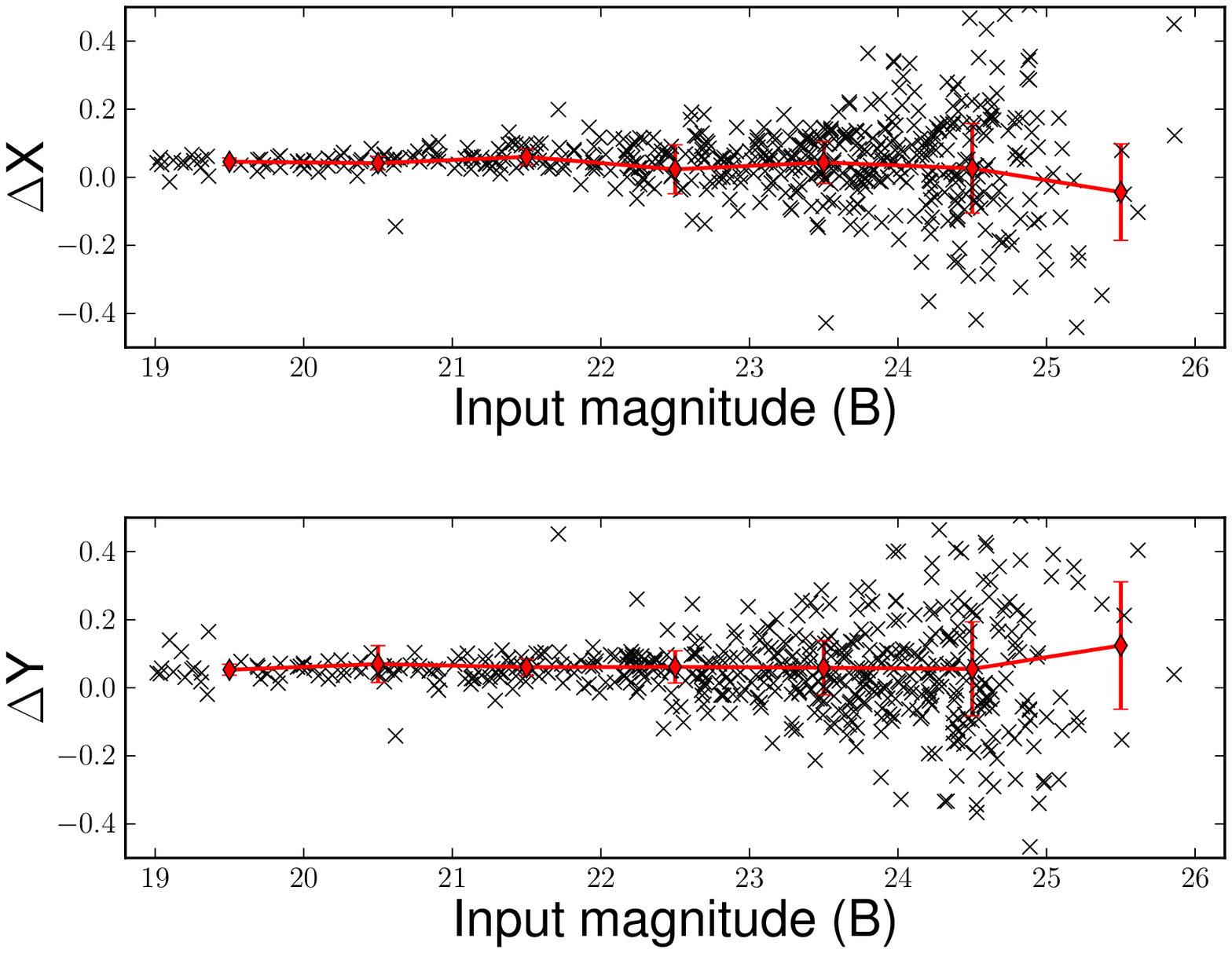}
 \label{fig8b}}
 \caption{Difference between barycenter coordinates estimated by DAOPHOT ({\it{left panels}}) and by SExtractor ({\it{right panels}}), and input coordinates and as a function of input magnitude for detected stars. Superimposed red points and solid red lines draw the mean and standard deviation values reported in top part of Tab.~\ref{tab5}.}
 \label{fig8}
 \end{figure*}

\begin{figure*}[!ht]
 \centering
 \subfloat[]{
 \includegraphics[scale=0.4]{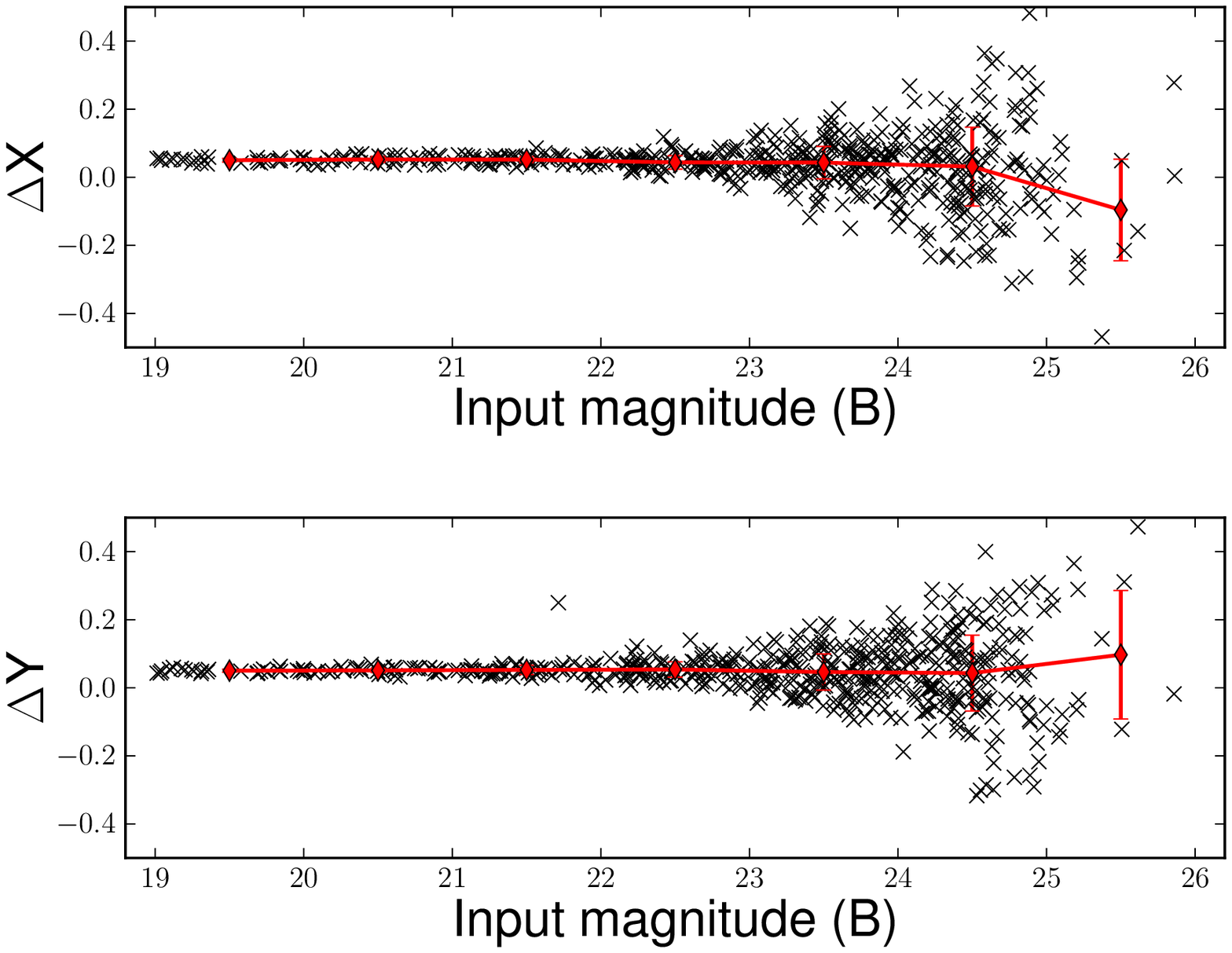}
 \label{fig9a}
 }
 \subfloat[]{
 \includegraphics[scale=0.4]{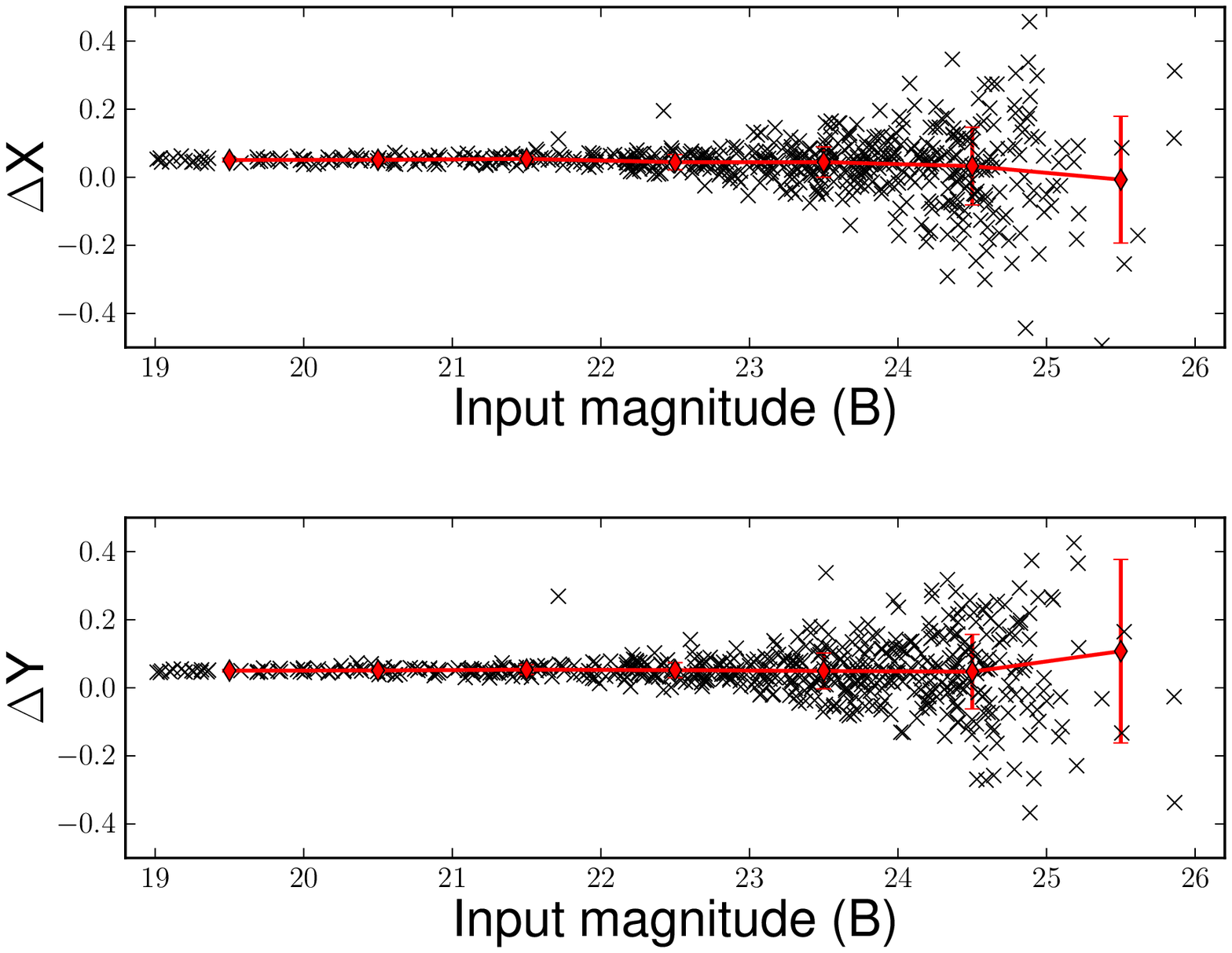}
 \label{fig9b}
 }
 \caption{Difference between PSF-corrected and input coordinates estimated by DAOPHOT ({\it{left panels}}) and by SExtractor ({\it{right panels}}), as a function of input magnitude for detected stars. Superimposed red points and solid red lines draw the mean and standard deviation values reported in bottom part of Tab.~\ref{tab5}.}
 \label{fig9}
 \end{figure*}

The last comparison is among extracted and input positions. There are different ways to obtain centroid measurements. As stated above, DAOPHOT can provide two different measurements for centroids. The simplest are the coordinates of the source barycenter, derived during the thresholding process. These coordinates can be redetermined by ALLSTAR, once DAOPHOT has built a PSF model, by applying a PSF correction.\\
Concerning SExtractor, we chose to compare the results obtained using the barycenter and the PSF corrected coordinates, as for DAOPHOT, and the results obtained by using the windowed positions along both axes. These coordinates are obtained by integrating pixel values within a circular Gaussian window.
In Tab.~\ref{tab5}, it is reported the mean difference between barycenter coordinates and PSF-corrected coordinates, estimated respectively with DAOPHOT (parts a and b) and SExtractor (parts c and d) and input coordinates.\\
Finally, Tab.~\ref{tab6} shows the difference among input and windowed coordinates estimated by SExtractor. \\
Fig.~\ref{fig8} and ~\ref{fig9} show the difference between the input and barycenter coordinates and between the input and PSF corrected coordinates.\\
Both software packages show a bias between output centroid coordinates $\le$ 0.01 arcsec (equal to $\sim$ 0.47 pixel) and input X and Y, with an average deviation of $\le$ 0.02 arcsec (equal to $\sim$ 0.94 pixel), down to the DAOPHOT completeness magnitude limit. These values are in particular improved in terms of average deviation ($\sigma_{\Delta X(Y)} \le$ 0.01 arcsec), when PSF correction is applied. Hence, we can conclude that the results for centroids are satisfactory in both cases.\\

 \begin{table*}[h]
\centering
\small
 \begin{tabular}{l c c || c c || c c || c c}
 \hline
 \hline
\rule[-1.0ex]{0pt}{1.0ex}  \tt{Bin} & \tt{$\mathrm{\Delta m_{mean}}$} &  \tt{$\mathrm{\sigma_{\Delta m}}$} & \tt{$\mathrm{\Delta m_{mean}}$} &  \tt{$\mathrm{\sigma_{\Delta m}}$} & \tt{$\mathrm{\Delta m_{mean}}$} &  \tt{$\mathrm{\sigma_{\Delta m}}$}& \tt{$\mathrm{\Delta m_{mean}}$} &  \tt{$\mathrm{\sigma_{\Delta m}}$} \\
\rule[-1.0ex]{0pt}{1.0ex}  (mag) & \tt{$\mathrm{(mag)}$} &  \tt{$\mathrm{(mag)}$} & \tt{$\mathrm{(mag)}$} &  \tt{$\mathrm{(mag)}$} & \tt{$\mathrm{(mag)}$} &  \tt{$\mathrm{(mag)}$} & \tt{$\mathrm{(mag)}$} & \tt{$\mathrm{(mag)}$}\\
\hline
\rule[-1.0ex]{0pt}{1.0ex}  & \multicolumn{2}{c ||}{\tt{$\mathrm{(a)}$}} & \multicolumn{2}{c ||}{\tt{$\mathrm{(b)}$}} & \multicolumn{2}{c ||}{\tt{$\mathrm{(c)}$}}  & \multicolumn{2}{c}{\tt{$\mathrm{(d)}$}}\\
\hline
\hline
 \rule[-1.0ex]{0pt}{1.0ex}  19 - 20 & 0.007 & 0.005 & 0.002 & 0.006 & -0.001 & 0.003 & 0.003 & 0.003 \\
 \rule[-1.0ex]{0pt}{1.0ex}  20 - 21 & 0.006  & 0.010 & -0.001 & 0.011 & -0.006  & 0.005 &  0.003  & 0.004 \\
 \rule[-1.0ex]{0pt}{1.0ex}  21 - 22 & 0.024 & 0.070 & -0.011 & 0.031 & -0.016 & 0.011 & 0.000 & 0.012 \\
 \rule[-1.0ex]{0pt}{1.0ex}  22 - 23 & -0.020 & 0.075 & -0.009 & 0.059 & -0.023 & 0.013 & 0.003 & 0.014 \\
 \rule[-1.0ex]{0pt}{1.0ex}  23 - 24 & -0.069 & 0.157 & -0.011 & 0.106 & -0.032 & 0.026 & 0.005 & 0.025 \\
 \rule[-1.0ex]{0pt}{1.0ex}  24 - 25 & -0.147 & 0.302 & -0.044 & 0.250 & -0.034 & 0.058 & -0.002 & 0.055 \\
 \rule[-1.0ex]{0pt}{1.0ex}  25 - 26 & -0.306 & 0.379 & -0.032 & 0.444 & -0.122 & 0.088 &  -0.129 & 0.129 \\
 \hline
 \hline
 \end{tabular}
\caption{The table reports, as a function of the magnitude bin (col.~1), the mean difference {\tt{$\mathrm{\Delta m_{mean}}$}} (columns~2, 4, 6 and 8), and the standard deviation {\tt{$\mathrm{\sigma_{\Delta m}}$}} (columns~3, 5, 7 and 9) between aperture and input magnitudes as estimated by DAOPHOT (part a) and by SExtractor (part b). Parts c and d report the mean difference and the standard deviation between PSF and input magnitudes as obtained by using DAOPHOT and SExtractor, respectively.}
\label{tab3}
\end{table*}

\begin{table*}[!ht]
\centering
\small
 \begin{tabular}{l c c || c c || c c}
 \hline
 \hline
\rule[-1.0ex]{0pt}{1.0ex}  \tt{Bin} & \tt{$\mathrm{\Delta m_{mean}}$} &  \tt{$\mathrm{\sigma_{\Delta m}}$} & \tt{$\mathrm{\Delta m_{mean}}$} &  \tt{$\mathrm{\sigma_{\Delta m}}$}  & \tt{$\mathrm{\Delta m_{mean}}$} &  \tt{$\mathrm{\sigma_{\Delta m}}$} \\
\rule[-1.0ex]{0pt}{1.0ex}  (mag) & \tt{$\mathrm{(mag)}$} &  \tt{$\mathrm{(mag)}$} & \tt{$\mathrm{(mag)}$} &  \tt{$\mathrm{(mag)}$} &  (mag) & \tt{$\mathrm{(mag)}$} \\
\hline
\rule[-1.0ex]{0pt}{1.0ex}  & \multicolumn{2}{c ||}{\tt{$\mathrm{(a)}$}} & \multicolumn{2}{c ||}{\tt{$\mathrm{(b)}$}} & \multicolumn{2}{c}{\tt{$\mathrm{(c)}$}}\\
\hline
\hline \rule[-1.0ex]{0pt}{2.5ex}  19 - 20 & 0.074 & 0.005 & 0.058 & 0.007 & 0.057 & 0.023\\
 \rule[-1.0ex]{0pt}{1.0ex}  20 - 21 & 0.077 & 0.009 & 0.073 & 0.012 & 0.048 & 0.025\\
 \rule[-1.0ex]{0pt}{1.0ex}  21 - 22 & 0.076 & 0.027 & 0.093 & 0.024 & 0.028 & 0.046\\
 \rule[-1.0ex]{0pt}{1.0ex}  22 - 23 & 0.076 & 0.046 & 0.128 & 0.039 & -0.002 & 0.074\\
 \rule[-1.0ex]{0pt}{1.0ex}  23 - 24 & 0.087 & 0.065 & 0.216 & 0.052 & -0.034 & 0.091\\
 \rule[-1.0ex]{0pt}{1.0ex}  24 - 25 & 0.051 & 0.143 & 0.389 & 0.121 & -0.098 & 0.146\\
 \rule[-1.0ex]{0pt}{1.0ex}  25 - 26 & 0.147 & 0.195 & 0.749 & 0.143 & -0.145 & 0.151\\
\hline
\hline
\end{tabular}
\caption{The table reports, as a function of the magnitude bin (col.~1), the mean difference {\tt{$\mathrm{\Delta m_{mean}}$}} (columns~2, 4 and 6), and the standard deviation {\tt{$\mathrm{\sigma_{\Delta m}}$}} (columns~3, 5 and 7) between Kron (part a), isophotal (part b), model (part c) and and input magnitudes as obtained by using SExtractor.}
\label{tab4}
\end{table*}

\begin{table*}
\centering
\small
 \begin{tabular}{l c c | c c || c c | c c}
 \hline
 \hline
\rule[-1.0ex]{0pt}{1.0ex}  \tt{Bin}
& \tt{$\mathrm{\Delta X_{mean}}$} &  \tt{$\mathrm{\sigma_{\Delta X}}$}
& \tt{$\mathrm{\Delta Y_{mean}}$} &  \tt{$\mathrm{\sigma_{\Delta Y}}$}
& \tt{$\mathrm{\Delta X_{mean}}$} &  \tt{$\mathrm{\sigma_{\Delta X}}$}
& \tt{$\mathrm{\Delta Y_{mean}}$} &  \tt{$\mathrm{\sigma_{\Delta Y}}$} \\
\rule[-1.0ex]{0pt}{1.0ex}  (mag)
& \tt{$\mathrm{(pixel)}$} &  \tt{$\mathrm{(pixel)}$}
& \tt{$\mathrm{(pixel)}$} &  \tt{$\mathrm{(pixel)}$}
& \tt{$\mathrm{(pixel)}$} &  \tt{$\mathrm{(pixel)}$}
& \tt{$\mathrm{(pixel)}$} &  \tt{$\mathrm{(pixel)}$} \\
\hline
\rule[-1.0ex]{0pt}{1.0ex}  & \multicolumn{4}{c ||}{\tt{$\mathrm{(a)}$}} & \multicolumn{4}{c }{\tt{$\mathrm{(b)}$}}\\
\hline
\hline
 \rule[-1.0ex]{0pt}{2.5ex}  19 - 20 & 0.033  & 0.055 & 0.063 & 0.062 &  0.046 & 0.011 & 0.053 & 0.016\\
 \rule[-1.0ex]{0pt}{1.0ex}  20 - 21 & 0.032  & 0.045 & 0.041 & 0.059 &  0.041 & 0.019 & 0.070 & 0.054\\
 \rule[-1.0ex]{0pt}{1.0ex}  21 - 22 & 0.058  & 0.061 & 0.043 & 0.050 &  0.060 & 0.024 & 0.061 & 0.024\\
 \rule[-1.0ex]{0pt}{1.0ex}  22 - 23 & 0.033  & 0.068 & 0.048 & 0.058 &  0.023 & 0.072 & 0.061 & 0.047\\
 \rule[-1.0ex]{0pt}{1.0ex}  23 - 24 & 0.046  & 0.108 & 0.043 & 0.090 &  0.043 & 0.062 & 0.059 & 0.079\\
 \rule[-1.0ex]{0pt}{1.0ex}  24 - 25 & 0.033  & 0.209 & 0.026 & 0.202 &  0.026 & 0.132 & 0.055 & 0.138\\
 \rule[-1.0ex]{0pt}{1.0ex}  25 - 26 & 0.027  & 0.167 & 0.107 & 0.288 & -0.044 & 0.142 & 0.124 & 0.187\\
\hline
\rule[-1.0ex]{0pt}{1.0ex}  & \multicolumn{4}{c ||}{\tt{$\mathrm{(c)}$}} & \multicolumn{4}{c }{\tt{$\mathrm{(d)}$}}\\
\hline
\hline
 \rule[-1.0ex]{0pt}{2.5ex}  19 - 20 &  0.050 & 0.005 & 0.050 & 0.004 & 0.051 & 0.005 & 0.050 & 0.003\\
 \rule[-1.0ex]{0pt}{1.0ex}  20 - 21 &  0.052 & 0.009 & 0.051 & 0.006 &  0.051 & 0.007 & 0.051 & 0.006\\
 \rule[-1.0ex]{0pt}{1.0ex}  21 - 22 &  0.052 & 0.009 & 0.053 & 0.014 &  0.054 & 0.009 & 0.053 & 0.014\\
 \rule[-1.0ex]{0pt}{1.0ex}  22 - 23 &  0.044 & 0.020 & 0.054 & 0.022 &  0.044 & 0.022 & 0.052 & 0.022\\
 \rule[-1.0ex]{0pt}{1.0ex}  23 - 24 &  0.043 & 0.048 & 0.046 & 0.053 &  0.044 & 0.044 & 0.049 & 0.052\\
 \rule[-1.0ex]{0pt}{1.0ex}  24 - 25 &  0.031 & 0.116 & 0.043 & 0.111 &  0.033 & 0.114 & 0.047 & 0.109\\
 \rule[-1.0ex]{0pt}{1.0ex}  25 - 26 & -0.096 & 0.149 & 0.097 & 0.189 & -0.007 & 0.186 & 0.107 & 0.269\\
\hline
\hline
\end{tabular}
\caption{The table reports, as a function of the magnitude bin (col.~1), the mean difference between DAOPHOT X (col.~2),Y (col.~4) barycenter measure and input X,Y with the relative standard deviation (columns~3 and 5) in the part a, while in the part b there are the mean difference between SExtractor X (col.~6),Y (col.~8) barycenter measure and input X, Y with the relative standard deviation (columns~7 and 9). In parts c and d are reported the mean difference between X (columns~2 and 6), Y (columns~4 and 8) PSF corrected and input measurements obtained by using DAOPHOT and SExtractor, respectively, with the relative standard deviation (columns~3, 5, 7 and 9).}
\label{tab5}
\end{table*}

\begin{table*}[h]
\centering
\small
 \begin{tabular}{l c c | c c}
 \hline
 \hline
\rule[-1.0ex]{0pt}{1.0ex}  \tt{Bin} & \tt{$\mathrm{\Delta X_{mean}}$} &  \tt{$\mathrm{\sigma_{\Delta X}}$} & \tt{$\mathrm{\Delta Y_{mean}}$} &  \tt{$\mathrm{\sigma_{\Delta Y}}$} \\
\rule[-1.0ex]{0pt}{1.0ex}  (mag) & \tt{$\mathrm{(pixel)}$} &  \tt{$\mathrm{(pixel)}$} & \tt{$\mathrm{(pixel)}$} &  \tt{$\mathrm{(pixel)}$} \\
\hline
\hline
 \rule[-1.0ex]{0pt}{2.5ex}  19 - 20 & 0.050 & 0.004 & 0.050 & 0.004\\
 \rule[-1.0ex]{0pt}{1.0ex}  20 - 21 & 0.051 & 0.006 & 0.051 & 0.007 \\
 \rule[-1.0ex]{0pt}{1.0ex}  21 - 22 & 0.054 & 0.010 & 0.056 & 0.016\\
 \rule[-1.0ex]{0pt}{1.0ex}  22 - 23 & 0.032 & 0.036 & 0.054 & 0.025\\
 \rule[-1.0ex]{0pt}{1.0ex}  23 - 24 & 0.044 & 0.046 & 0.052 & 0.057\\
 \rule[-1.0ex]{0pt}{1.0ex}  24 - 25 & 0.028 & 0.127 & 0.040 & 0.120\\
 \rule[-1.0ex]{0pt}{1.0ex}  25 - 26 & -0.043 & 0.124 & 0.134 & 0.173\\
\hline
\hline
\end{tabular}
\caption{The table reports, as a function of the magnitude bin (col.~1), the mean difference between X (col.~2) and Y (col.~4) windowed and input measurements as estimated by SExtractor with the relative standard deviation (columns~3 and 5).}
\label{tab6}
\end{table*}

 \clearpage
 \twocolumn
\section{Non uniform star distribution}
\label{6}
\begin{figure*}[!ht]
\centering
\subfloat{
\includegraphics[height=6 cm, width=8 cm ]{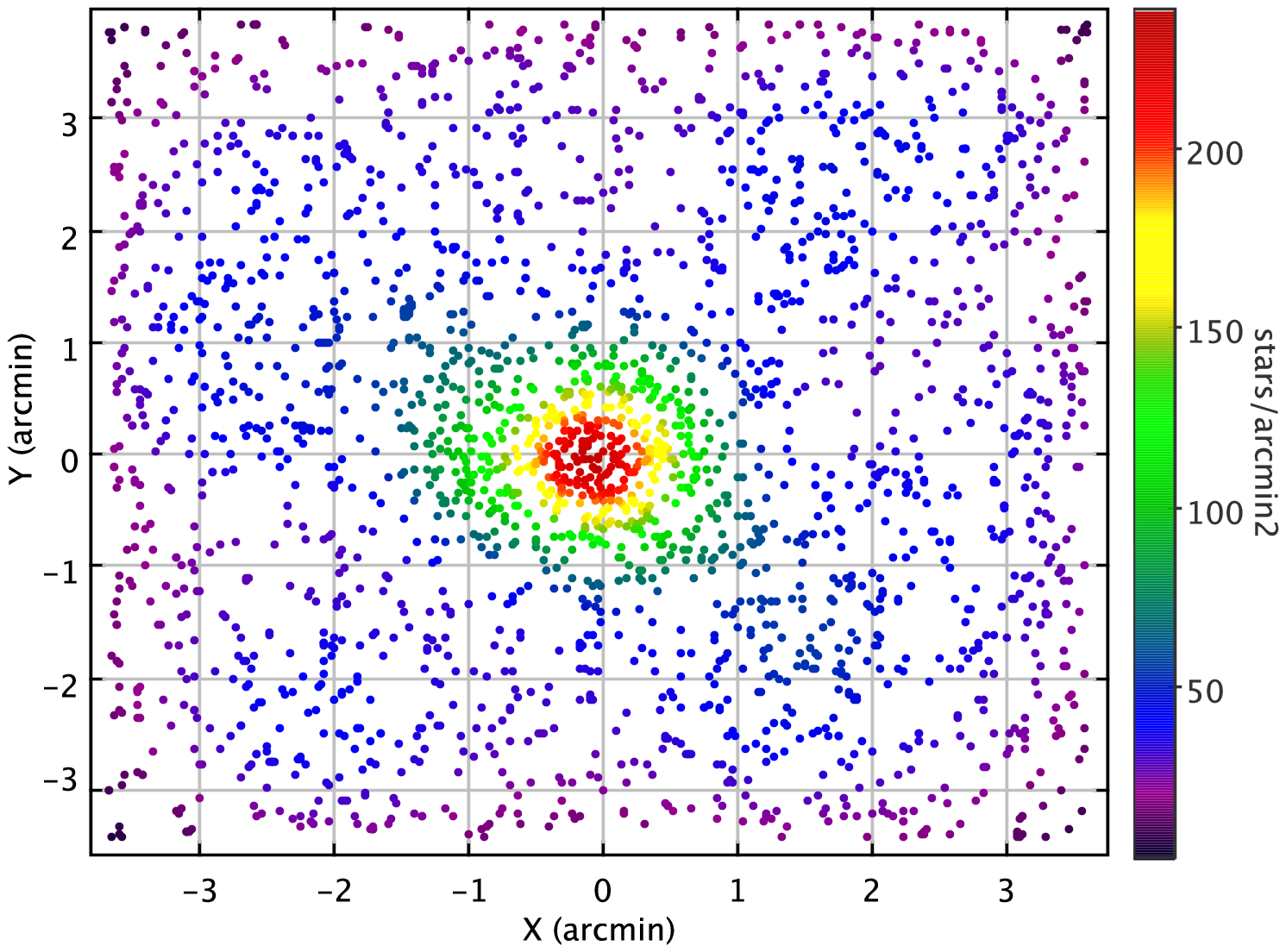}
\label{fig10a}}
\subfloat{
\includegraphics[height=6 cm, width=8 cm]{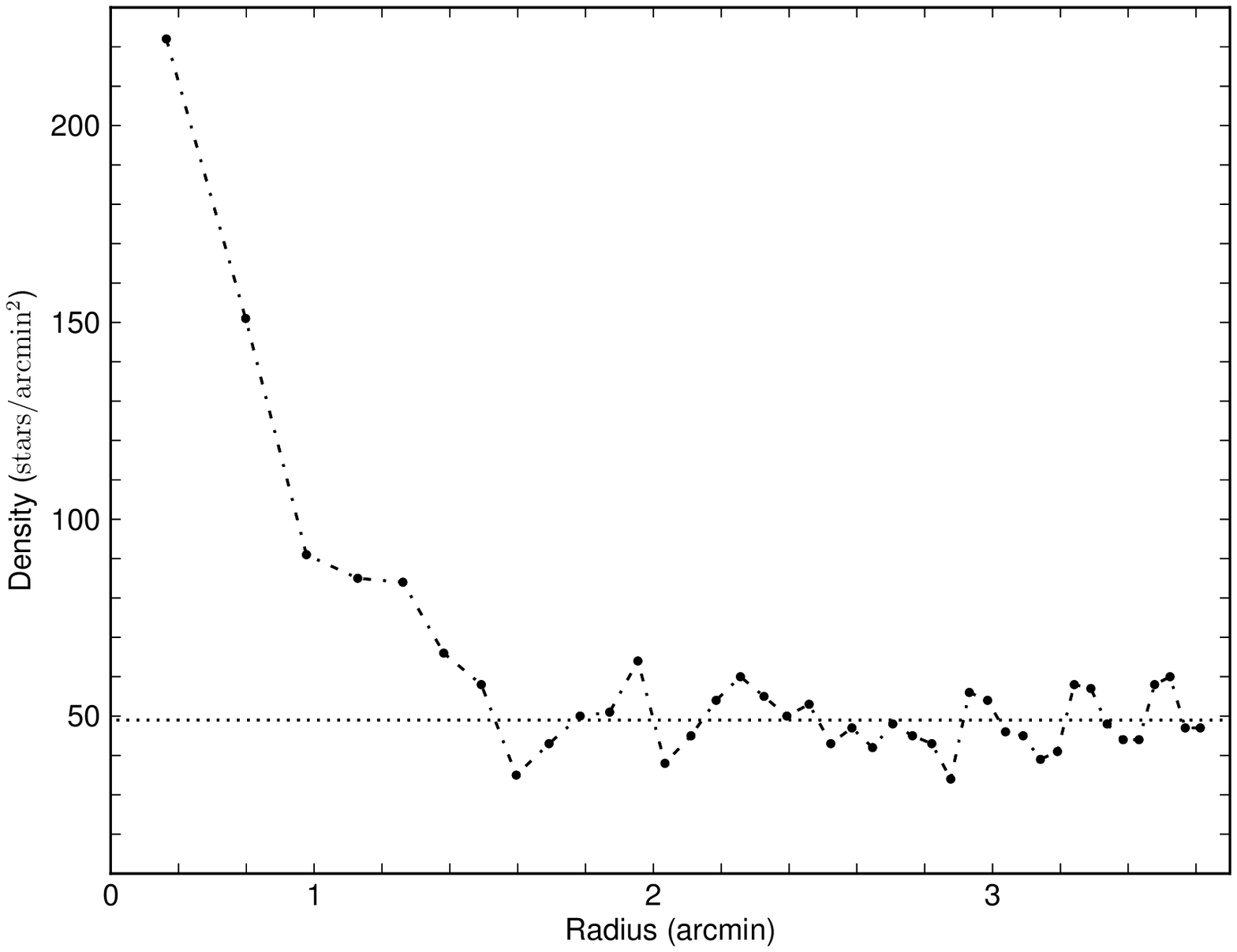}
\label{fig10b}}
\caption{\textit{Left panel}: Spatial distribution of input stars for the over-dense stellar field. Colors of points refer to the surface number density of stars. Density values are reported in the vertical color-bar. \textit{Right panel}: Numerical stellar density as a function of the distance from the center. Dotted line represents the average stellar density ($\sim$ 40 stars/arcmin$\mathrm{^2}$) obtained excluding the circular region within a radius of 1.5 arcmin from the center. }
\label{fig10}
\end{figure*}

In order to evaluate the performances of both software in crowded fields, we tested them on a simulated image with a non uniform stellar distribution, i.e., showing an overdensity of stars in the center. In the left panel of Fig.~\ref{fig10}, we plot the spatial distribution of the simulated stars, while the stellar density of the field, as a function of the distance from the center, is shown in the right panel. 

In order to carry out the test, we fixed the values for DAOPHOT, ALLSTAR, SExtractor and PSFEx as in Sect.s~\ref{4.1} and \ref{4.2}.\\
The obtained results were evaluated by following the same criteria as in Sect.~\ref{5}. For all tests, we excluded saturated stars, which are stars with magnitude brighter than B = 17 mag.\\
Also, in case of a non uniform stellar field, we obtain that, for point-like sources, the final depth returned by DAOPHOT is $\sim$ 2 mag brighter than those produced by SExtractor. In fact, SExtractor can recover the input star catalog, resulting in a photometric depth of B = 25 mag, while for DAOPHOT the resulted limit in magnitude is around 23 mag.\\
Moreover, by considering the reliability of the extracted catalogs, it is possible to separate stars and galaxies down to the magnitude limit for each software package: B = 25 mag for SExtractor and B = 23 mag for DAOPHOT.\\
In order to not bias the comparison of photometric measurements, we consider only the input stellar sources recovered by both software packages.
As a first step, we analyze results for aperture photometry. Comparing the values of the mean and the standard deviation of the difference between aperture and input magnitudes with those reported in Tab.~\ref{tab3}, we obtained a decline in the quality of the results, in particular, those produced by SExtractor. In fact, in the magnitude bin 22-23, we go from $\mathrm{\Delta m} \pm \sigma_{\Delta m}$ = -0.009 $\pm$ 0.059 mag in the case of non-crowded field, to -0.165$\pm$ 0.287 mag for SExtractor and from $\mathrm{\Delta m} \pm \sigma_{\Delta m}$ -0.020 = 0.075 $\pm$ mag to -0.487 $\pm$ 0.474 for DAOPHOT.\\
By using PSF-fitting photometry, there is an improvement in the determination of magnitude with respect to aperture photometry, but the results show a quality decline as compared to the case of less crowded stars.\\ 
By considering the last bin, for DAOPHOT PSF-fitting photometry, the results change from $\mathrm{\Delta m} \pm \sigma_{\Delta m}$ = -0.023 $\pm$ 0.013 mag to -0.079 $\pm$ 0.080 mag. While, by using SExtractor PSF-fitting photometry, we pass from $\mathrm{\Delta m} \pm \sigma_{\Delta m}$ = -0.003 $\pm$ 0.014 mag to -0.063 $\pm$ 0.110 mag. \\
Finally, we compare the results among those obtained with barycenter and PSF corrected centroids and those reported in Tab.~\ref{tab5}.
In the last bin of magnitude, DAOPHOT measurements of PSF centroids go from $\mathrm{\Delta X} \pm \sigma_{\Delta X}$; $\mathrm{\Delta Y} \pm \sigma_{\Delta Y}$ = -0.044 $\pm$ 0.020;  -0.054 $\pm$ 0.022 pixel to $\mathrm{\Delta X} \pm \sigma_{\Delta X}$; $\mathrm{\Delta Y} \pm \sigma_{\Delta Y}$ = -0.045 $\pm$ 0.146;  -0.040 $\pm$ 0.148 pixel.\\
On the other hand, in the same bin, SExtractor measurements of PSF centroids go from $\mathrm{\Delta X} \pm \sigma_{\Delta X}$; $\mathrm{\Delta Y} \pm \sigma_{\Delta Y}$ = -0.044 $\pm$ 0.022;  -0.052 $\pm$ 0.022 pixel to $\mathrm{\Delta X} \pm \sigma_{\Delta X}$; $\mathrm{\Delta Y} \pm \sigma_{\Delta Y}$ = -0.063 $\pm$ 0.133;  -0.055 $\pm$ 0.150 pixel.\\
On the basis of this comparison, we can conclude that, as shown in Sect~\ref{5}, the results obtained by both packages, in the case of an overdense region, are completely equivalent both for PSF fitting photometry and PSF fitting determination of centroids, and with regard of the reliability of the extracted catalog. With SExtractor, however, we can optimize the detection of the sources to recover the whole input catalog. 

\section{Summary and conclusions}
\label{7}

The advent of new photometric surveys and the need to deal with fainter sources have led to an increase in the demand for quality and accuracy of photometric measurements. Moreover, the analysis of massive data sets needs a single software tool in order to minimize the number of required reprocessing steps.
Therefore, a crucial aspect in this context is the choice of source detection software package. An important innovation was introduced since 2010 with the public release of PSFEx, which can work in combination with SExtractor. This new software package has deeply improved the performances of SExtractor, filling the gap against other available procedures, such as DAOPHOT and PHOTO, concerning PSF photometry. \\
In the present work, for the first time, the performances of DAOPHOT and SExtractor are compared to check the quality and accuracy of extraction procedures with respect to the requirements of modern wide surveys. The software packages are tested on two kinds of B-band simulated images having, respectively, a uniform spatial distribution of sources and an overdensity in the center.\\
In the first case, by considering only the number of extracted sources, it appears that the limiting magnitude for the extracted catalog is extremely low, in particular for DAOPHOT is B = 22 mag. If we limit to consider only stellar sources, the photometric depth is improved down to 24 mag for DAOPHOT and 26 mag for SExtractor. This could be related to the fact that only SExtractor gives the possibility to choose the filter for catalog extraction, according to the image characteristics (see Sect.~\ref{4.2}). In fact, as shown in Sect~\ref{5.1}, the filter choice affects the photometric depth of the extracted catalog.\\
A relevant aspect of the catalog extraction is the capability to discriminate between extended and point-like sources. As we have seen, within the different software packages, there are various methods to perform the star/galaxy classification. In particular the sharpness parameter available in DAOPHOT, improved by using ALLSTAR, returns a reliable star/galaxy classification down to the photometric depth of the catalog (B = 24 mag). All the traditional methods available in SExtractor, instead, limit the star/galaxy classification, at least one magnitude above the completeness magnitude of the catalog. The new parameter $\mathrm{\tt{SPREAD\_MODEL}}$, which is a discriminant between the best fitting local PSF and a more extended model, has largely improved the classification, allowing to separate extended and point-like sources down to the completeness limit of the catalog (B = 26 mag for stellar sources).\\
Since DAOPHOT is mainly designed to perform stellar photometry, in order to not bias the comparison of photometric measurements, we considered only input stellar sources recovered by both software packages.
Both software tools are able to deliver acceptable performances in both aperture (with a $\mathrm{\sigma_{\Delta m}}\, <$ 0.2 mag) and PSF photometry (with a $\mathrm{\sigma_{\Delta m}}\, <$ 0.03 mag), down to B = 24 mag, the completeness limit of the DAOPHOT catalog.\\
Moreover, since SExtractor allows us to derive different estimates of the total magnitudes of sources, which we can also compare among themselves: Kron, isophotal and model magnitudes. The isophotal magnitude is highly dependent on the detection threshold. In fact, in the [23-24] magnitude bin, there is a higher shift of $\mathrm{\Delta m}$ (0.216 mag) than in other magnitudes. The Kron magnitude yields $\sim$ 94\% of the total source flux within the adaptive aperture (\citealp{bertin96}). Accordingly, we find a shift of $\sim$ 0.07 mag even in the brightest magnitude bin. The model magnitude results a good estimate of the input magnitude also for stars, with an error of 0.091 mag in the [23 - 24] magnitude bin.  \\
An accurate determination of the object's centroids is crucial for the relative astrometry and thus, also for matching sources in different bands. Both software packages show a bias between output centroids and input X and Y coordinates $\le$ 0.01 arcsec, with an average deviation of $\le$ 0.02 arcsec down to B = 24 mag. These values are improved in terms of average deviation ($\sigma_{\Delta X(Y)} \le$ 0.01 arcsec), when PSF correction is applied. So, we can conclude that these results are satisfactory in both cases.\\
We also tested both software packages on simulated images with a non uniform stellar distribution, i.e., showing an overdensity of stars in the center. By analyzing the results obtained in this last case, we can confirm the conclusions described above. In particular, DAOPHOT provides a catalog $\sim$ 2 mag shallower than the one extracted by SExtractor.
On the other hand, by analyzing the extracted catalogs in terms of the mean difference and the standard deviation among output and input magnitudes and centroids, we notice a decline in the quality of the results for both software packages, with respect to the case of a uniform spatial distribution of stars.
Finally, DAOPHOT and ALLSTAR provide very accurate and reliable PSF photometry, with a robust star-galaxy separation. However, it is not useful for galaxy characterization.
On the other hand SExtractor, associated with PSFEx, turns competitive in terms of PSF photometry. It returns acceptable aperture photometry and accurate PSF modeling also for faint sources. The windowed centroids are as good as PSF centroids. Moreover, SExtractor allows to go very deep in source detection through a properly choice of image filtering masks. The deblending model is very extensible, and the use of neural networks for object classification, plus the novel {\tt{SPREAD\_MODEL}} parameter, push down to the limiting magnitude the capability of star/galaxy separation. Considering that SExtractor returns accurate photometry also for galaxies, we can conclude that the new version of SExtractor, used in combination with PSFEx, represents a very powerful software tool for source extraction, with performances comparable to DAOPHOT also for overdense stellar fields. In the next years, it will be important to extensively test SExtractor plus PSFEx on real crowded stellar fields in order to definitively assess the performances of this software tool. However, an important aspect for the use of PSFEx and SExtractor, that we cannot avoid to mention, is the processing time. Without considering problems, such as degradation in performances during periods of heavy disk access, on average, SExtractor requires 0.5 $s$ per detection to perform PSF photometry and source modeling, by using one single CPU with 6GB of RAM. This suggests that currently, the only disadvantage of using SExtractor and PSFEx on wide field images is the processing time. However, on the other hand, although DAOPHOT is more efficient in terms of processing time just for the calculation, it requires more time if the user visually inspects the modeled PSF stars.

\section*{Acknowledgments}
Authors wish to thank the referee R. Laher for his positive comments and suggestions.
The present work has been carried out by an intensive exploitation of software packages DAOPHOT II with ALLSTAR, and SExtractor with PSFEx. The authors wish to thank E. Bert\`{\i}n for useful and helpful suggestions during the work, mainly related on the new unstable versions of the software. AM and MB wish to thank the financial support of PRIN-INAF 2010, Architecture and Tomography of Galaxy Clusters. The authors also wish to thank the financial support of Project F.A.R.O. III Tornata (P.I. Dr. M. Paolillo, University Federico II of Naples). MB acknowledges the financial support from the PRIN MIUR 2009: Gamma Ray Bursts: dai progenitori alla Fisica del processo dell'emissione ``prompt".


\end{document}